\documentclass[journal=jctcce,manuscript=article]{achemso}
\usepackage{graphics,graphicx,amsfonts,amsmath,amsbsy,amssymb,color}
\usepackage{verbatim}  
\usepackage{psfrag}  
\usepackage{tabularx}
\usepackage{xmpmulti}
\usepackage{hyperref}
\usepackage{marginnote}
\usepackage{subfigure}
\usepackage{paracol}
\usepackage{xparse}
\usepackage{layouts}
\usepackage{afterpage}
\usepackage{braket}
\usepackage{paralist}
\usepackage{bm}
\usepackage{url}
\usepackage[normalem]{ulem}
\usepackage{color}
\usepackage[colorinlistoftodos]{todonotes}
\usepackage[american]{babel} 
\usepackage[version=3]{mhchem}

\newcommand{\refeq}[1]{{Eq.~(\ref{#1})}}
\newcommand{\reffig}[1]{{Fig.~\ref{#1}}}

\newcommand{\iDMQMC}{\ensuremath{i-}DMQMC\ }

\title{Using density matrix quantum Monte Carlo for calculating exact-on-average energies for \emph{ab-initio} Hamiltonians in a finite basis set}

\author{Hayley~R.~Petras}
\affiliation[First University]
{Department of Chemistry, University of Iowa}
\alsoaffiliation[Another]
{University of Iowa Informatics Initiative, University of Iowa}

\author{Sai~Kumar~Ramadugu}
\affiliation[First University]
{Department of Chemistry, University of Iowa}
\alsoaffiliation[Another]
{University of Iowa Informatics Initiative, University of Iowa}

\author{Fionn~D.~Malone}
\affiliation[Second University]{Quantum Simulations Group, Lawrence Livermore National Laboratory, 7000 East Avenue, Livermore, CA, 94551 USA.}

\author{James~J.~Shepherd}
\email{james-shepherd@uiowa.edu}
\affiliation[First University]
{Department of Chemistry, University of Iowa}
\alsoaffiliation[Another]
{University of Iowa Informatics Initiative, University of Iowa}
\date{\today}

\begin{document}

\begin{abstract}

We here apply the recently developed initiator density matrix quantum Monte Carlo (i-DMQMC) 
to a wide range of chemical environments using atoms and molecules in vacuum. i-DMQMC samples the exact density matrix of a Hamiltonian at finite temperature and combines the accuracy of full configuration interaction quantum Monte Carlo (FCIQMC) – full configuration interaction (FCI) or exact energies in a finite basis set – with finite temperature. By way of exploring the applicability of i-DMQMC for molecular systems, we choose to study a recently developed test set by Rubenstein and coworkers: \ce{Be}, \ce{H2O}, and \ce{H10} at near-equilibrium and stretched geometries. 
We find that, for \ce{Be} and \ce{H2O}, i-DMQMC delivers energies which are sub-millihartree accuracy when compared with finite temperature FCI. 
For \ce{H2O} and both geometries of \ce{H10} we examine the difference between FT-AFQMC and i-DMQMC which in turn is an estimate of the difference in canonical versus grand canonical energies. 
We close with a discussion of simulation parameters (initiator error and different basis sets) and by showing energy difference calculations in the form of specific heat capacity and ionization potential calculations.
\end{abstract}
\date{\today}
\maketitle

\section{Introduction}

The last decade has seen a remarkable growth in the study or warm dense matter.\citep{graziani_frontiers_2014,dornheim_review}
Encompassing conditions intermediate between plasma and condensed matter physics, warm dense matter is relevant to the study of planetary interiors\citep{benuzzi-mounaix_progress_2014,koenig_progress_2005}, hot-electron chemistry\citep{mukherjee_hot_2013-1} and laser excited solids\citep{ernstorfer_formation_2009}.
Experimentally, warm dense matter can now be routinely investigated in large scale laser facilities\citep{glenzer_slac_15}.
However, the theoretical description of warm dense matter is challenging due to equal importance of electron-electron, quantum degeneracy and thermal electronic effects.
Finite temperature density functional theory (FT-DFT)\citep{mermin_tdft}, coupled with molecular dynamics, is currently the method of choice, due to its relatively low computational cost and often satisfactory agreement with experiment\citep{zhang_hydrocarbons_17}.
However, unlike its ground state counterpart, there is not the same level of understanding of its limitations and dependence on the form of exchange correlation free energy functional used\citep{karasiev_importance_2016,karasiev_tlda_14,groth_fxc_17,karasiev_xc_deuterium}.
Thus, there is a need to develop more accurate approaches which can benchmark, and potentially supplement FT-DFT.

Much like the more familiar zero temperature wavefunction-based quantum chemistry methods there is in principle a hierarchy of finite temperature methods starting with thermal Hartree--Fock theory\citep{mermin_thf} and ending in finite temperature full configuration interaction theory\citep{kou_finite-temperature_2014}.
The performance of this hierarchy of approaches in a quantum chemical context have only recently begun to be explored.
For example,  finite temperature perturbation theories\cite{he_finite-temperature_2014,hermes_finite-temperature_2015}, Green's function methods \cite{rusakov_self-consistent_2016}  as well as coupled cluster theory \cite{hummel_finite_2018,white_time-dependent_2018} have all been developed in recent years with promising results.
Other approaches include thermofield theory\cite{harsha_thermofield_2019} with extensions to time-dependent coupled cluster\citep{harsha_thermofield_2019-1} and cumulant-based approaches\cite{sanyal_thermal_1992}.

Quantum Monte Carlo (QMC) methods offer an alternative stochastic approach to simulating systems at finite temperature.
QMC methods are attractive as they offer a favourable scaling with system size $\mathcal{O}(N^3-N^4)$ and can scale on modern supercomputing architectures.
Unfortunately, they also suffer from the fermion sign problem at low temperatures and large system sizes that can only be overcome at polynomial cost by imposing a constraint. The bias resulting from this constraint, while typically small, can only be systematically removed at an exponential cost in general.

Of the finite temperature QMC methods available, real space path integral Monte Carlo (PIMC) is perhaps the most widely used\citep{ceperley_pimc}. It has the significant advantage of working in the complete basis set limit which is otherwise challenging to reach at finite temperature due to the thermal occupation of virtual states.
As a result, PIMC is already a standard method in the simulation of warm dense matter\citep{driver_all-electron_2012,driver_wdm_oxygen}. To overcome the sign problem at low temperature, one can use the restricted path integral formalism (RPIMC)\citep{ceperley_fermion_nodes}, similar in spirit to the fixed-node approximation in diffusion Monte Carlo (DMC)\citep{foulkes_quantum_2001}, which enforces a constraint using a trial density matrix. The quality of this constraint is largely unknown, however results for the uniform electron gas suggest it is unreliable at high densities and at lower temperatures\citep{brown_path-integral_2013,schoof_textitab_2015,malone_accurate_2016}.
We note there have been promising developments in extending the scope of PIMC to lower temperatures and higher densities through algorithmic developments\citep{dornheim_pbpimc,dubois_sign,schoof_cpimc}.

Auxiliary-Field QMC (AFQMC) is another promising QMC method capable of simulating matter at finite temperature\citep{blankenbecler_dqmc_1,scalapino_dqmc,zhang_ftafqmc_99,rubenstein_finite-temperature_2012}. In contrast to PIMC, AFQMC works in a second quantized framework, utilizing the Hubbard-Stratonovich transformation to write the partition function as an integral over auxiliary fields of fermion determinants. Motivated by the remarkable accuracy of AFQMC for the ground state properties of model Hamiltonians and ab-initio systems\citep{simons_hubbard_2d,motta_towards_2017,motta_review}, Liu \emph{et al.} recently extended the finite temperature phaseless AFQMC algorithm\citep{zhang_phaseless} to simulate ab-initio systems and developed a set of molecular benchmarks for small atoms and molecules for which no constraint was required.\cite{liu_ab_2018}
Each of these systems had its temperature-dependent internal energy calculated in the grand canonical ensemble, in a vacuum, and also in a finite basis set: Be (MIDI) atom, \ce{H2O} (STO-3G) molecule, \ce{C2} (STO-6G) molecule, \ce{H10} (STO-6G), and stretched \ce{H10} (STO-6G). This interesting test set deserves attention because it represents a range of different chemical environments, incorporating both weak and strong correlation.

In this paper we investigate the ability of an alternative QMC method, the density matrix quantum Monte Carlo method (DMQMC)\citep{blunt_density-matrix_2014} in its initiator variant\cite{malone_accurate_2016}  (i-DMQMC) to simulate real ab-initio systems.
DMQMC is the finite temperature analogue of the full configuration interaction QMC method (FCIQMC)\citep{booth_fermion_2009} and is capable of simulating systems outside the reach of conventional FCI without imposing a constraint.
To date, DMQMC has been applied to simulate the 2D-Heisenberg model\citep{blunt_density-matrix_2014} and the warm dense uniform electron gas\citep{malone_interaction_2015,malone_accurate_2016}, however its performance for real systems is largely unknown.
DMQMC is a promising tool in the benchmarking of finite temperature methods as it provides access to a statistical representation of the $N$-electron thermal density matrix. Thus, arbitrary expectation values can be evaluated as well as free energy differences\citep{malone_accurate_2016} and Renyi entropies \citep{blunt_density-matrix_2014} which are often a challenge for QMC methods.
Moreover, it can build on the many advancements made in the FCIQMC community such as the initiator approximation\citep{cleland_communications:_2010}, semi-stochastic approaches\cite{petruzielo_semistochastic_2012, blunt_semi-stochastic_2015}, better excitation generators \cite{holmes_efficient_2016, li_fast_2018}, and perturbative corrections \citep{ blunt_communication:_2018, blunt_preconditioning_2019}.  Additional exciting advancements based on coupled cluster equations include cluster-analysis-driven FCIQMC (CAD-FCIQMC) for accelerating convergence of FCIQMC, \cite{deustua_communication:_2018} as well as the EOMCC(P) approach, that uses the early stages of FCIQMC to converge excited state energies at the EOMCCSD(T) level of theory.\cite{deustua_accurate_2019}. Stochastic approaches that build off of the  deterministic CC(P;Q) methodology have also been developed.  \cite{shen_combining_2012, bauman_combining_2017, deustua_converging_2017} 

Given that the scope of benchmark systems accessible to FCIQMC includes the UEG, as well as atoms, ions, molecules, dimers, and most recently, solids\citep{booth_towards_2013}, it is important to investigate to what extent DMQMC follow this success. To begin answering this question, we present a set of calculations on previously established benchmark systems.

Our paper is organized as follows: In Section 2, we present a brief overview of the DMQMC method, following the derivation from Blunt and coworkers\citep{blunt_density-matrix_2014} and including the initiator approximation to DMQMC. Section 3 presents the comparison of full configuration interaction to i-DMQMC, a comparison to FT-AFQMC, a demonstration of the initiator approximation in DMQMC, i-DMQMC results for \ce{H2O} in three finite basis-sets and finally, additional applications of finite-temperature results. In Section 4, we conclude with final thoughts and future work.  %

\section{Methods}

Following the derivation from Blunt and coworkers,  \cite{blunt_density-matrix_2014} it can be shown that the $N$-particle density matrix in the canonical ensemble can be sampled by solving the symmetrized Bloch equation using a stochastic approach.
Starting with the thermal density matrix
\begin{equation}
\hat{\rho}(\beta)=e^{-\beta\hat{H}},
\end{equation}
where $\hat{H}$ is the Hamiltonian operator and $\beta=\frac{1}{k_{B}T}$  is the inverse temperature, differentiation with respect to $\beta$ shows that the density matrix obeys
the symmetrized Bloch equation  
\begin{equation}
\frac{d\hat{\rho}(\beta)}{d\beta}=-\frac{1}{2}(\hat{H}\hat{\rho}+\hat{\rho}\hat{H})=-\frac{1}{2} \{ {\hat{H},\hat{\rho}} \}.
\end{equation}
The density matrix at any temperature can then be found using a finite difference approach:
\begin{equation}
\hat{\rho}(\beta+\Delta\beta)=\hat{\rho}(\beta)-\frac{\Delta\beta}{2}(\hat{H}\hat{\rho}(\beta)+\hat{\rho}(\beta)\hat{H})+O(\Delta\beta^2)\label{eq:fd}
\end{equation}
where a finite time step $\Delta\beta$ has been introduced.
\refeq{eq:fd} coupled with the the initial condition $\hat{\rho}(\beta=0)=\hat{I}$ is sufficient to determine to  density matrix at any temperature.

To proceed we represent the density matrix in a basis of outer products of Slater determinants and rewrite \refeq{eq:fd} as
\begin{equation}
\rho_{\mathbf{ij}}(\beta+\Delta\beta)= \rho_{\mathbf{ij}}(\beta) + \Delta \rho_{\mathbf{ij}}(\beta) 
\end{equation}
\begin{equation}
\begin{split}
\Delta \rho_{\mathbf{ij}}(\beta) &= - \frac{\Delta\beta}{2}\displaystyle\sum_{\mathbf{k}}[(H_{\mathbf{ik}}-S\delta_{\mathbf{ik}})\rho_{\mathbf{kj}}-\rho_{\mathbf{ik}}(H_{\mathbf{kj}}-S\delta_{\mathbf{kj}})] \\
&=\frac{\Delta\beta}{2}\displaystyle\sum_{\mathbf{k}}(T_{\mathbf{ik}}\rho_{\mathbf{kj}}+\rho_{\mathbf{ik}}T_{\mathbf{kj}}).
\end{split}\label{eq:iter},
\end{equation}
where $\rho_{\mathbf{ij}}=\langle D_{\mathbf{i}}|\hat{\rho}|D_{\mathbf j} \rangle$  and $|D_{\mathbf i}\rangle$ is a Slater determinant, 
$S$ is a variable shift to be defined, and $T_{\mathbf{ij}}=-(H_{\mathbf{ij}}-S\delta_{\mathbf{ij}})$.
The idea of DMQMC is to now introduce a population of $N_w$ signed walkers which sample elements of the density matrix and evolve according to \eqref{eq:iter}. 
At $\beta=0$ walkers are uniformly distributed among the diagonal density matrix elements.
At each time step walkers then undergo a series of spawning cloning/death and annihilation steps analogous to those in FCIQMC.
Spawning is the probability a particle will spawn from a density matrix element  $\rho_{\mathbf {ik}}$ to $\rho_{\mathbf {ij}}$ and is given as $p_{s}(\mathbf{ik} \to \mathbf{ij})=\frac{\Delta\beta |T_{\mathbf{kj}}|}{2}$, with $\text{sign}(\rho_{\mathbf{ij}}) = \text{sign}(\rho_{\mathbf {ik}}) \times \text{sign}(T_{\mathbf{kj}})$, with a similar expression for $p_{s}(\mathbf{kj}\to \mathbf{ij})$.
The cloning/death process, where the number of walkers on a given density matrix element is either increased or decreased, is given by probability of $p_{d}(\mathbf{ij})=\frac{\Delta\beta}{2}|T_{\mathbf{ii}} + T_{\mathbf{jj}}|$. The particles clone if $\text{sign}(T_{\mathbf{ii}}+T_{\mathbf{jj}}) \times \text{sign}(\rho_{\mathbf {ij}}) > 0$ and die otherwise.
 To help control the sign problem, particles of opposite signs on the same density matrix element are annihilated; this annihilation does not affect the distribution of particles.
Similar to other QMC methods we must use population control to avoid either a population explosion or collapse.
We use a variable shift, S, which is adjusted according to
\begin{equation}
S(\beta+A\Delta\beta)=S(\beta)-\frac{\zeta}{A\Delta\beta}\ln\Big(\frac{N_{w}(\beta)+A\Delta\beta}{N_{w}(\beta)}\Big)
\end{equation}
where A is the number of beta steps between shift updates, $\zeta$ is a shift damping parameter, $N_{w}(\beta)$ is the total number of walkers at $\beta$. The shift damping parameter is chosen by the user when necessary to prevent large fluctuation in the shift.
This process is repeated until a desired inverse temperature is reached.
One complete evolution of walkers is called a $\beta$-loop and we average results over many such (independent) $\beta$-loops to obtain statistical estimates for physical observables.

At a specific temperature, the quantum mechanical expectation value for any quantum mechanical operator, $\hat{O}$, can be found through the stochastic sampling of 
\begin{equation}
\langle \hat{O} \rangle = \frac{\mathrm{Tr}(\hat{\rho} \hat{O})}{\mathrm{Tr}(\hat{\rho})}=\frac{\frac{1}{N_\beta}\sum_{\alpha}^{N_\beta} \sum_{\mathbf{ij}} w^{(\alpha)}_{\mathbf{ij}} O_{\mathbf{ji}}}{\frac{1}{N_\beta}\sum_{\alpha}^{N_\beta}\sum_{\mathbf{i}} w^{(\alpha)}_{\mathbf{ii}}},
\end{equation}
where $w_{\mathbf{ij}}^{(\alpha)}$ is the walker population on $\rho_{\mathbf{ij}}$ for simulation $\alpha$ of $N_\beta$ $\beta$-loops and the $\beta$ dependence has been omitted from the expression for clarity.
The numerator and denominator are sampled separately over the course of the evolution,  and the average for each value is calculated at discrete temperature values. Calculations are averaged over a number of $\beta$-loops until the desired statistical accuracy is achieved.

The goal of the initiator approximation for DMQMC (i-DMQMC)\cite{malone_accurate_2016} is similar to that of i-FCIQMC\cite{cleland_initiator_2009}: Restrict the spawning of walkers from negligibly  small elements to other negligibly small elements. A threshold is set to determine "initiator determinants", where the walker population on those elements are higher than that of the threshold.  The approximation then restricts the algorithm such that only the "initiator determinants" can spawn children onto unoccupied matrix elements. Children can also be spawned from multiple sign-coherent events. In the infinite population limit, the initiator approximation is recovered as the orginal DMQMC algorithm. 

The DMQMC
calculations that follow used a timestep of 0.001 (Ha$^{-1}$).  The target population was chosen to be $5 \times 10^6$; we show in section IIIA that the finite-temperature results are converged at a target population of $10^6$, similar to FCIQMC. We found that 25 $\beta$-loops gave us the desired statistical accuracy without being overly computationally demanding. Each calculation completed in less than a week. 
All of the calculations that follow were performed with the initiator approximation to DMQMC (\iDMQMC)\citep{malone_accurate_2016}, unless noted otherwise.
All calculations were performed using the HANDE code\citep{spencer_hande-qmc_2019}.

\section{Results}
To assess the capabilities of i-DMQMC on small molecules, we have performed a series of simulations on a small set of systems.  We first present a comparison to finite temperature and ground state full configuration interaction (FCI) for \ce{Be} and \ce{H2O} to assess the accuracy of the method, and show that \iDMQMC can treat small molecules and atoms without  modifications such as importance sampling\cite{blunt_density-matrix_2014} or the interaction picture\cite{malone_interaction_2015}.  We then provide a comparison of \ce{H2O} in the canonical (\iDMQMC) and grand canonical ensembles (FT-AFQMC). Next, we investigate the rate of convergence of \iDMQMC with respect to target population and show \iDMQMC results for \ce{H2O} in a varity of basis-sets to determine the ease of applicability of \iDMQMC to molecular systems.  Finally, due to the small errors in the \iDMQMC results, we investigate additional applications of \iDMQMC, where we show that we can take energy differences and calculate properties such as specific heat and ionization energy as functions of temperature.

\begin{figure}
\begin{center}
\subfigure[\mbox{}]{%
\includegraphics[width=0.4\textwidth,height=\textheight,keepaspectratio]{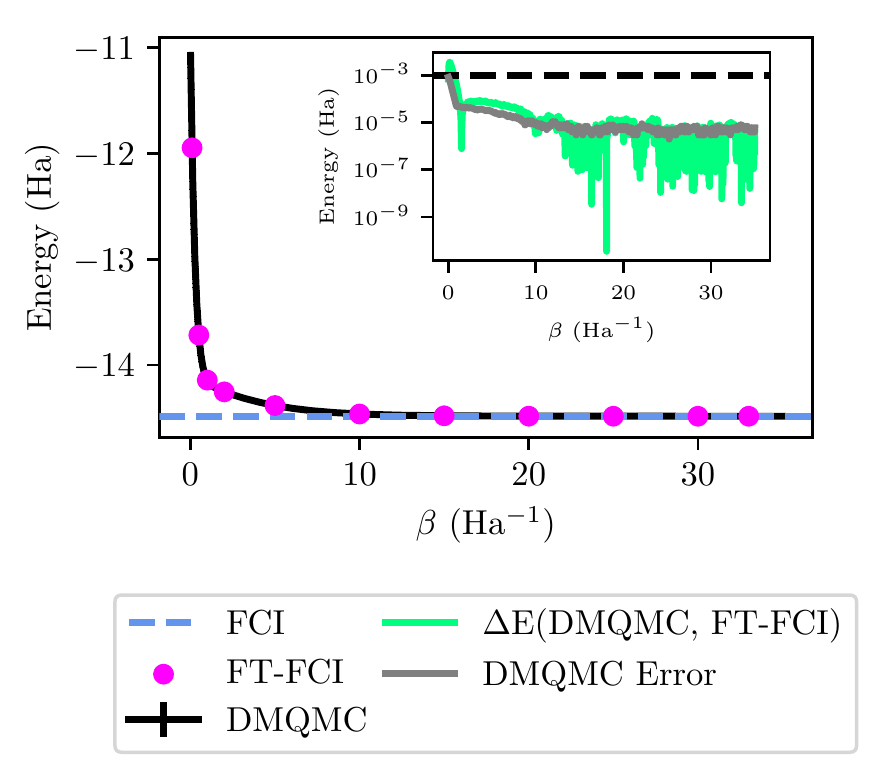}}
\subfigure[\mbox{}]{%
\includegraphics[width=0.4\textwidth,height=\textheight,keepaspectratio]{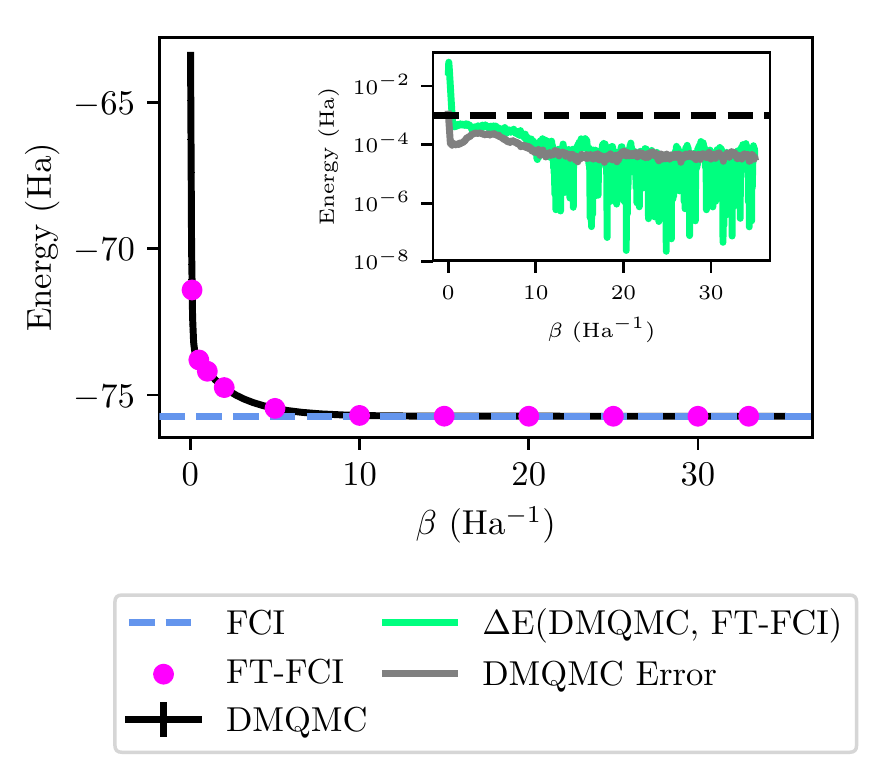}
\label{h2o_fci}}

\caption{The  energy of the (a) \ce{Be} atom (MIDI) and (b) \ce{H2O} (STO-6G) at various inverse temperatures using \iDMQMC (black), FT-FCI (magenta) and ground state FCI (blue). The integral files were generated in pySCF\cite{sun_pyscf:_2018}; \ce{H2O} was generated with an \ce{O-H} bond length of 0.96\AA\, and a bond angle of 109.5 degrees. The error bars are plotted in the main graph, but may be too low to see their actual scale. The inset presents the statistical errors (grey) more clearly, 
as well as the absolute difference between FT-FCI and i-DMQMC (green),
plotted on a logarithmic y-axis. As a reference, a black dashed line is provided in the inset at 10$^{-3}$.   FCI results were obtained using the FCI routine in HANDE-QMC; FT-FCI is described in the text. The \iDMQMC simulation was run with a timestep of 0.001 Ha$^{-1}$, the target population was set to $5 \times 10^6$, and run to $\beta$= 35. }
\label{FCI}
\end{center}
\end{figure}

\begin{figure}
    \centering
\includegraphics[width=0.4\textwidth,height=\textheight,keepaspectratio]{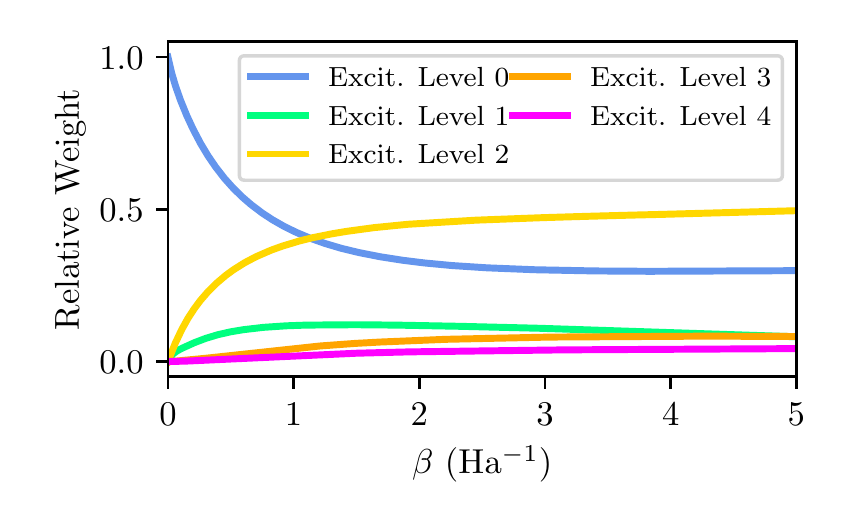}
    \caption{The relative weights of matrix elements in a single $\beta$-loop simulation of \ce{H2O} (STO-6G).   The \iDMQMC simulation was run with a timestep of 0.001 Ha$^{-1}$, the target population was set to $5 \times 10^6$, and run to $\beta$= 5. The higher the relative weight in, the more off-diagonal the element is.}
    \label{h2o_weights}
\end{figure}

\begin{figure}
    \centering
    \includegraphics[width=0.5\textwidth,height=\textheight,keepaspectratio]{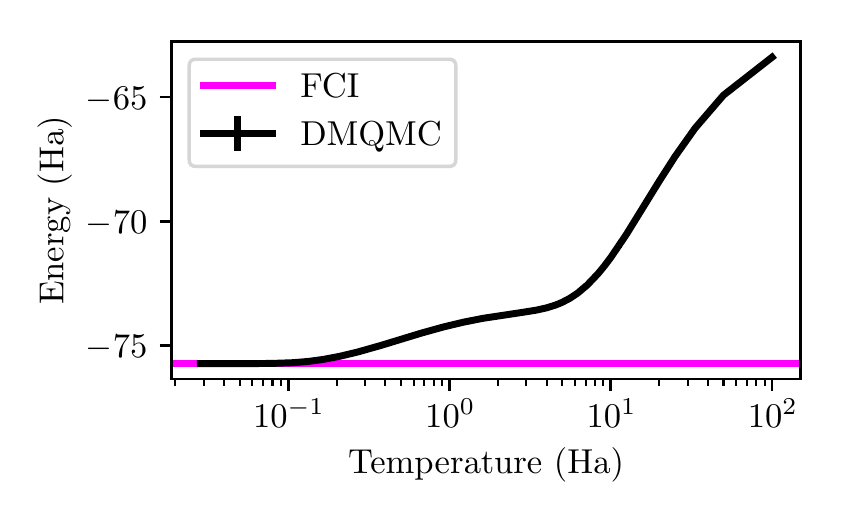}
    \caption{The i-DMQMC results for \ce{H2O} (STO-3G, black) as a function of \textit{temperature}, plotted on a logarithmic x-axis, with the ground state FCI value (magenta) shown for comparison. The statistical error is plotted for the i-DMQMC result, but is invisible on this scale. }
    \label{zerotemp}
\end{figure}

\subsection{Agreement between i-DMQMC and FCI}

In this section we first compare our \iDMQMC results on small molecules to FT-FCI.
For such small systems, we can determine all eigenstates of the Hamiltonian and compute the temperature-dependent internal energy ($U$) using: $\langle \hat{H} \rangle = \frac{\sum_i E_i e^{-\beta E_i }} {\sum_i e^{-\beta E_i }} $, where $E_i$ is an eigenvalue of the FCI Hamiltonian computed in a given basis set.

In \reffig{FCI},  \iDMQMC is compared with FT-FCI for  (a) the beryllium atom in the MIDI basis set and (b) \ce{H2O} in the STO-6G basis set.   The energies were calculated for a range of inverse temperatures ($0.01 \leq \beta \leq 35.00$) and averaged over 25 $\beta$-loops. %
Both i-DMQMC and FT-FCI can be computed straightforwardly for these systems so the sampling in $\beta$ can be made arbitrarily fine. 
As can be seen in both figures, the FT-FCI points agree with the \iDMQMC calculations to within the \iDMQMC statistical error bar ($\pm\sigma_{\bar{U}}$), indicating that the systematic error has been eliminated to at least below the statistical error. \iDMQMC and ground state FCI values agree within $2\sigma$ at $\beta>5$. At $\beta<5$, the deviations are larger but the energy gap is still small relative to the total energy.
The statistical error bar in the \iDMQMC calculation is invisible on the graphs as shown but has been added as an inset. The errors in both the \ce{Be} and \ce{H2O} systems are sub-milliHartree, and a dashed line has been provided on the insets at 10$^{-3}$ as a reference. The \iDMQMC and FCI errors are the same order of magnitude, and although the difference is greater than the error at some points, we attribute this to an error in the error calculation.

We used 25 $\beta$ loops for these simulations in order to achieve error bars of the desired accuracy
; this is a significantly lower number of loops 
than previous studies but we wanted to consider what was a typical cost-effective simulation.\cite{blunt_density-matrix_2014} The time taken to run the \iDMQMC simulation on \ce{H2O} to $\beta =$ 35 over 25 $\beta$-loops is 110 core-hours. 

The relative weights for each state in a single $\beta$-loop  simulation on \ce{H2O} are shown in \reffig{h2o_weights}, for $0.00 < \beta \leq 5.00$ where the higher the relative weight, the more off diagonal the element is. It can be seen from this plot that the relative weight of the diagonal (excitation level 0) dominates the simulation at $\beta<1$. The off-diagonal elements, besides excitation level 2,  have low relative weights, showing the sparsity of the density matrix. This is significant because analyzing the distribution of weights in the density matrix can provide information for other methods that utilize trial density matrices, such as AFQMC.

In \reffig{zerotemp}, we show the \iDMQMC energy for \ce{H2O} as  a function of \emph{temperature}, compared to the ground state FCI energy. We expect the \iDMQMC energy to converge to the FCI ground state in the zero-temperature limit, and this is confirmed in the plot. This is a critical test for finite-temperature methods to assess the accuracy of the method because if the simulation does not converge to the ground state FCI energy in the zero temperature limit, then this is indicative of a sampling error. Our prior work investigated this in the context of the uniform electron gas.\cite{malone_accurate_2016}  
As the temperature approaches zero, the energy we obtain using \iDMQMC approaches the FCI energy which is seen both in this figure and in \reffig{h2o_fci}, where the \iDMQMC result overlaps with the ground state FCI.
At temperatures lower than 0.1 Ha, the statistical error in \iDMQMC makes it impossible to distinguish the energy difference from zero. These results are noteworthy in light of the fact that we have not yet applied algorithmic improvements beyond the initiator approximation; in future studies it would be interesting to see the effect of importance sampling\cite{blunt_density-matrix_2014} or the interaction picture.\cite{malone_interaction_2015} By showing agreement between \iDMQMC and ground state FCI at the zero temperature limit, as well as agreement between FT-FCI and the \iDMQMC results, we have shown evidence that strongly suggests \iDMQMC is a highly accurate method for treating small molecular and atomic systems. 
\subsection{Comparison to AFQMC}

 \begin{figure}
    \centering
    \subfigure[\mbox{}]{
    \includegraphics[width=0.5\textwidth,height=\textheight,keepaspectratio]{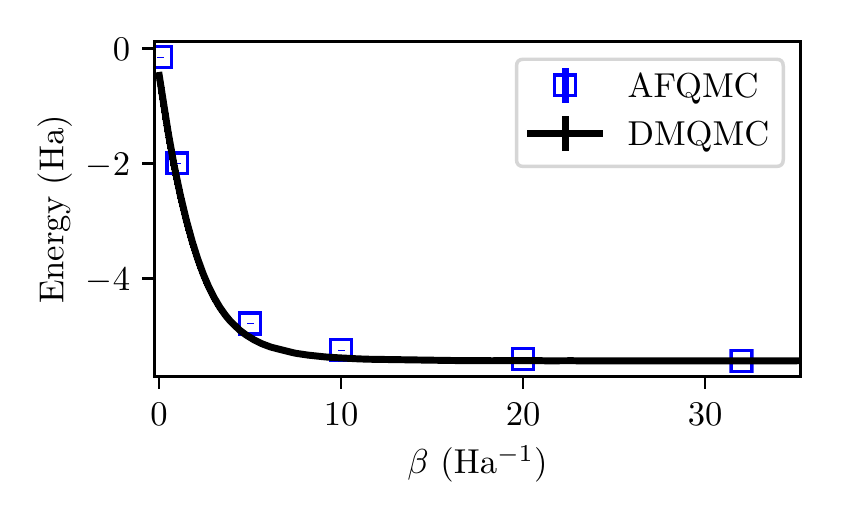}}
    \subfigure[\mbox{}]{
    \includegraphics[width=0.5\textwidth,height=\textheight,keepaspectratio]{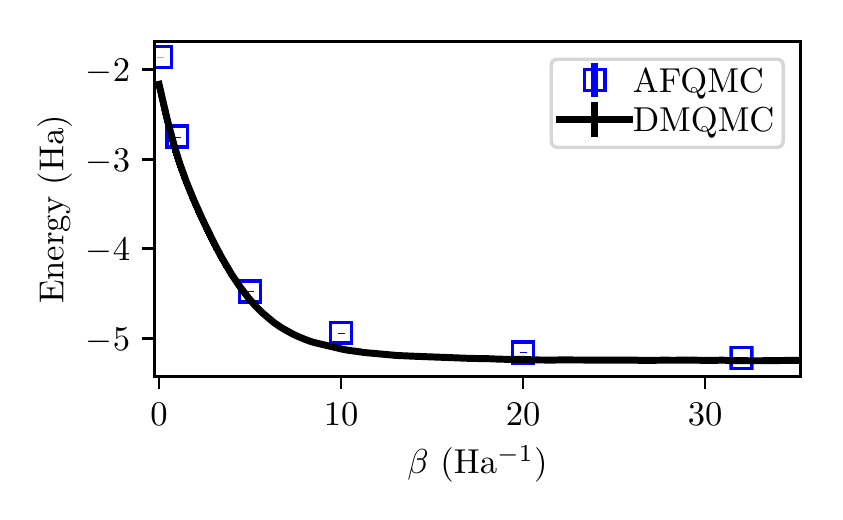}}
    \subfigure[\mbox{}]{
    \includegraphics[width=0.5\textwidth,height=\textheight,keepaspectratio]{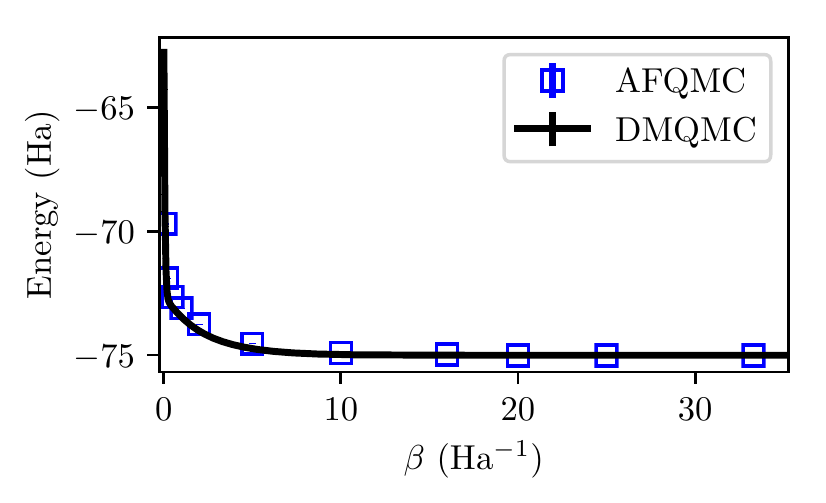}}
    \caption{The \iDMQMC results of (a) equilibrium \ce{H10},  (b) stretched \ce{H10} and (c) \ce{H2O} (STO-3G) compared to FT-AFQMC at discrete values of $\beta$. The equilibrium \ce{H10} has a bond length of 1.786 \AA and the stretched \ce{H10} molecule has a bond length of 2.4 \AA and was generated using the STO-6G basis set.  The integral files for the \ce{H10} systems were generated in MOLPRO \cite{MOLPRO-WIREs}.} 
    
    \label{afqmc}
\end{figure}

 \begin{figure}
    \centering
    \includegraphics[width=0.5\textwidth,height=\textheight,keepaspectratio]{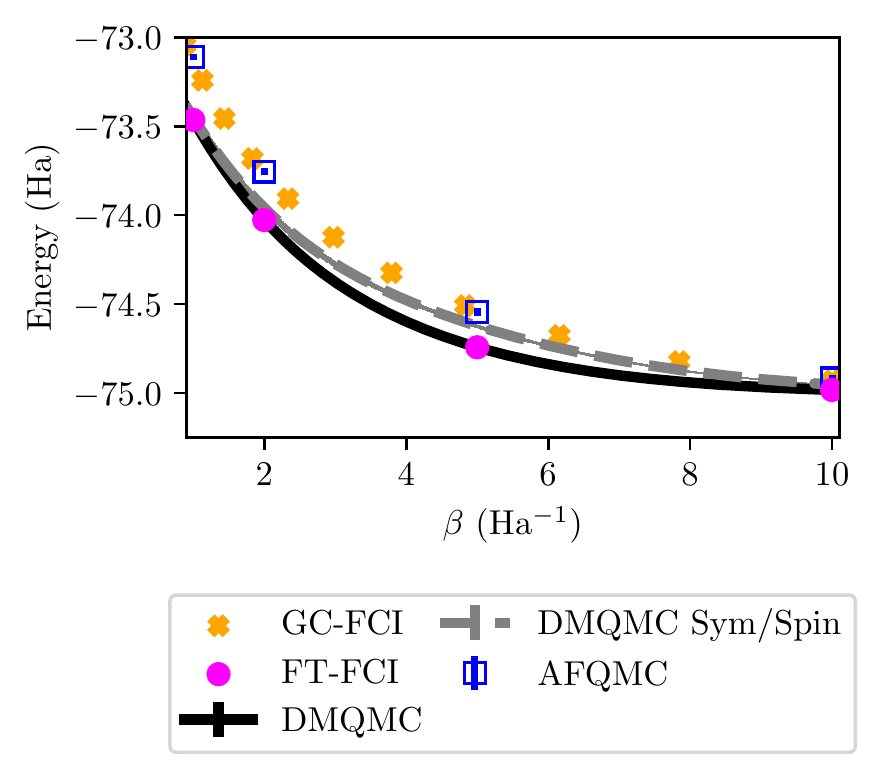}
    \caption{The energy results of three different methods:\iDMQMC (black), \iDMQMC, sampled over all symmetries and spin polarizations (grey, dotted); GC-FCI (orange crosses); FT-FCI (magenta circles) and FT-AFQMC(blue squares). The same \ce{H2O} geometry was used for these calculations as above, but the integral was produced in the STO-3G basis set for comparison with Rubenstein and coworkers. The \iDMQMC calculation details can be found in the main text.} 
    \label{GCFCI_AFQMC}
\end{figure}

Due to the success of the recently developed finite-temperature auxiliary field quantum Monte Carlo (FT-AFQMC) method for small molecules, atoms, solids and models, we choose to directly compare our results for an equilibrium and stretched \ce{H10} molecule (STO-6G) and \ce{H2O} (STO-3G) to Rubenstein and coworkers.\cite{liu_ab_2018}
One significant difference between the \iDMQMC method utilized here and the FT-AFQMC used by Rubenstein and coworkers is that they are performed in different ensembles. FT-AFQMC works in the grand canonical ensemble, while i-DMQMC works in the canonical ensemble. It is possible for both of these methods to work in each of the two ensembles, but this is not addressed here. 
We note that differences between the two ensembles are expected to be most pronounced in the small systems studied here. 

We present FT-AFQMC results, as reported by Rubenstein and coworkers,\cite{liu_ab_2018} in comparison with \iDMQMC results in \reffig{afqmc} for \ce{H10}, stretched \ce{H10} and \ce{H2O} (STO-3G) in the range of $0.01 \leq \beta \leq 35.00$. 
The difference between the FT-AFQMC and the \iDMQMC results cannot be seen clearly in this figure, because the energy differences are small on this scale. The greatest differences are at low values of $beta$ for each of the three systems. For \ce{H2O} the highest difference occurs at 0.1 Ha$^{-1}$, where the absolute difference between the two methods is approximately 1.01 Ha. The difference between the two methods decreases as $\beta$ increases, which suggests they are both converging to the zero-temperature limit.  

In \reffig{GCFCI_AFQMC} we have plotted both grand canonical FCI (GC-FCI) and FT-FCI results with the FT-AFQMC and \iDMQMC results for an intermediate $\beta$ range. 
  The FT-AFQMC agrees with the GC-FCI results, as is to be expected, while the FT-FCI results agree with the \iDMQMC simulations.  

To investigate whether the differences between the \iDMQMC and FT-AFQMC results is indeed an artifact of ensemble differences, the internal energy of \ce{H2O} was also calculated in \reffig{GCFCI_AFQMC}  wherein all spin sectors are sampled over (DMQMC Sym/Spin), as opposed to the usual case when only elements with $M_s=0$ are sampled. 
As can be seen from the figure, this approximation is well controlled with the effect of spin sampling being most noticeable in intermediate $\beta$ ranges. 

\subsection{Initiator error and basis sets}

\begin{figure}
\includegraphics[width=0.6\textwidth,height=\textheight,keepaspectratio]{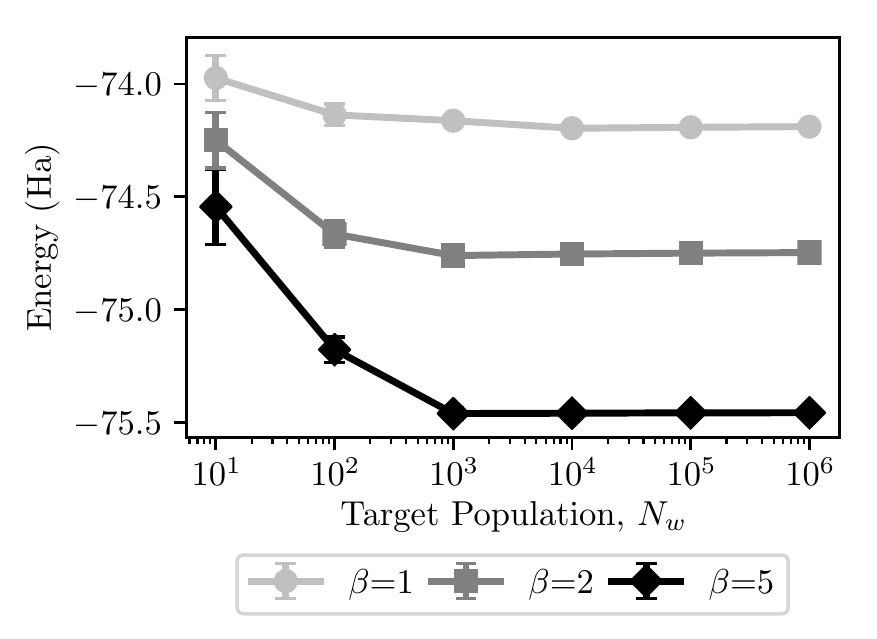}
\caption{ The \iDMQMC energies of \ce{H2O} (STO-6G) at $\beta = $1 (circles), $\beta = $2 (squares) and $\beta = $5 (diamonds), as a function of target population. \iDMQMC was performed with 6 different target populations of 10 to $10^6$ in logarithmic steps.The error bars are plotted for each marker, but may be invisible on this scale for the larger target populations.  }
\label{h2o_fciqmc} 
\end{figure}

\begin{figure}
    \centering
    
\includegraphics[width=0.6\textwidth,height=\textheight,keepaspectratio]{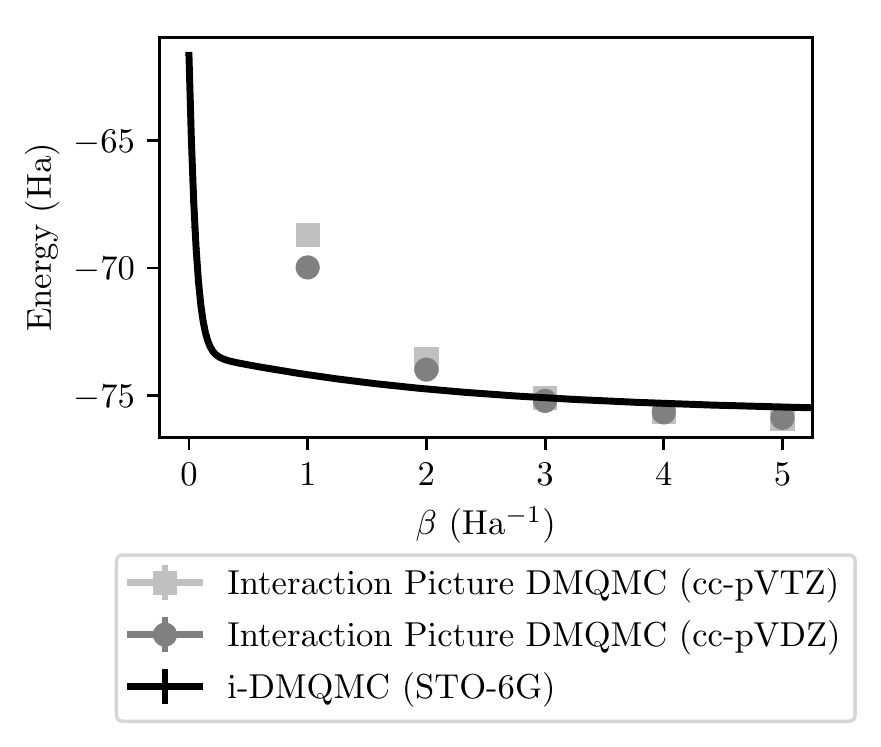}
    \caption{Energy results for $0.01 \leq \beta \leq 5.00$ for \ce{H2O} in three basis sets: STO-6G (black), cc-pVDZ (grey circles) and cc-pVTZ (light grey squares)  The number of orbitals in each basis set for \ce{H2O} is 7, 24 and 58, respectively. .  The results for cc-pVDZ and cc-pVTZ were obtained using the interaction picture variant of DMQMC \cite{malone_interaction_2015}.   The simulations were run over 25 $\beta$-loops. The interaction picture calculations were performed with a timestep of 0.0001 Ha$^{-1}$ and a target population of $10^5$.}
    \label{h2o_basissets}
\end{figure}

The initiator approximation is used in i-DMQMC to exploit the sparsity of the density matrix by controlling the ability of walkers to spawn from very small matrix elements to other small elements,\cite{malone_accurate_2016} as adapted from the FCIQMC initiator method.\cite{cleland_initiator_2009} The initiator approximation was designed to restrict spawning from negligibly small matrix elements to other small matrix elements. This is accomplished by setting a $n_{add}$, a threshold that determines a set of "initiator determinants", where particles can only be spawned to unoccupied matrix elements if the particles originate from the set of elements with a population greater than $n_{add}$. Here,  $n_{add}$ = 3.0. 

In Figure \ref{h2o_fciqmc} we compare the rate of convergence for three values of $\beta$ with respect to target population for \iDMQMC.   The energy decreases with increasing target population until the energy is converged. As shown in the plot, the results for $\beta$=2 and $\beta$=5 converge at a target population of $10^3$. The absolute differences between $N_w$=$10^3$ and $N_w$=$10^4$ for $\beta$=2 and $\beta$=5 are 0.007 Ha and 0.001 Ha respectively. $\beta$=1 converges at a slightly higher population of $10^4$, and the absolute difference between $N_w$=$10^4$ and $N_w$=$10^5$ is 0.004 Ha. For this target population range, the error is visible for $N_w$=$10$ and $N_w$=$10^2$, but decreases as $N_w$ increases, such that the error bar becomes invisible on this scale.  We can see from the relatively fast convergence rate and decreasing error, that \iDMQMC is well-controlled with respect to target population. This data suggests that the \iDMQMC energy converges similarily to i-FCIQMC with respect to target population, and is a vital part of showing the validity of this method in the long-term.  

\reffig{h2o_basissets} compares the \iDMQMC results of \ce{H2O} in the STO-6G basis set compared to the interaction picture \iDMQMC results in the cc-pVDZ and cc-pVTZ basis sets. 
The interaction picture is a modification to the original DMQMC algorithm that was developed to stabilize sampling in larger basis sets at the cost of only sampling one $\beta$ at a time.\cite{malone_interaction_2015} 
We see that the internal energy converges with basis set differently depending on the $\beta$ value. 
At high temperatures the internal energy increases which is caused by the increase in kinetic energy as higher single-particle states become occupied. As the temperature is decreased (large $\beta$) we see that increasing the basis set size serves to lower the internal energy, which is to be expected as we approach the $T=0$ limit. Comparing the three basis-sets, we see that \iDMQMC controls the error for all three relatively well. This is similar to FCIQMC and shows promise for the applicability of \iDMQMC to molecular systems. The use of interaction picture DMQMC has not yet been studied extensively; that work is out of the scope of this paper, but is a future line of inquiry. 

\subsection{Applications with energy differences}

\begin{figure}
    \centering
    \includegraphics[width=0.5\textwidth,height=\textheight,keepaspectratio]{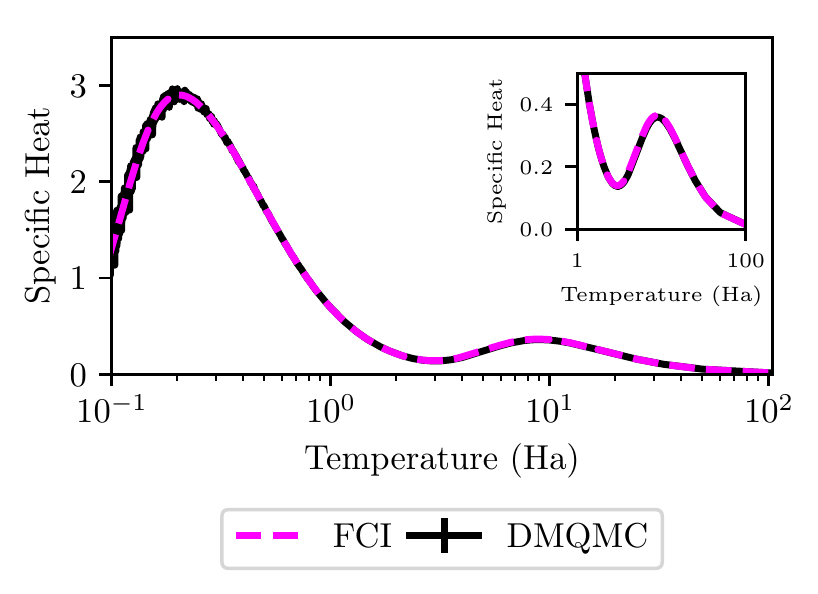}
    \caption{  The specific heat of \ce{H2O}, calculated numerically as described in the text, from the internal energy output of the \iDMQMC calculation (black) and from FT-FCI (magenta).  Note the specific heat presented is dimensionless because DMQMC works in atomic units. The specific heat is presented as a function of temperature, with temperature on a logarithmic scale in order to see the low temperature limit  with more detail. The propagated error is plotted for the \iDMQMC result, but is invisible on this scale.} 
    \label{specheat_h2o}
\end{figure}

The internal energy is not the only physical observable that can be studied from the i-DMQMC method.
 We demonstrate here that due to the small errors, this method is a strong candidate for taking energy differences, such as those that are used for calculating numerical derivatives and those used for calculating the ionization energy of atoms and molecules.  We show the specific heat as a function of temperature can be calculated from \iDMQMC with  accuracy at the FT-FCI level.

The specific heat, $C_V$, was computed from the usual expression
\begin{equation}
    C_V = \left(\frac{\partial U}{\partial T}\right)_{V} = -\frac{1}{k_B T^2}\left(\frac{\partial U}{\partial \beta}\right)_V,
\end{equation}
where we used finite-differences to numerically compute the derivative of the internal energy. 
The results for the specific heat of \ce{H2O} in the STO-6G basis-set
, are shown in \reffig{specheat_h2o} for both \iDMQMC and FT-FCI. By visual inspection, the curves overlap exactly, which is indicative of the two methods being in agreement. The specific heat of \ce{H2O} in two larger basis sets, cc-pVDZ and cc-pVTZ, were also investigated, and the results of these plots showed a similar shape to the STO-6G data, but the noise is overwhelming at higher temperature values than that of the STO-6G results. 
At lower temperature this procedure became overwhelmed by noise, which is to be expected due to issues differentiating stochastic functions\citep{tpoole_2015}. 
A more direct approach, would be to compute $\frac{1}{k_B T^2}(\langle \hat{H}^2 \rangle -\langle \hat{H} \rangle^2)$ directly in the simulation, however this would be computationally challenging and we did not attempt it here.

\begin{figure}
    \centering
    \includegraphics[width=0.6\textwidth,height=\textheight,keepaspectratio]{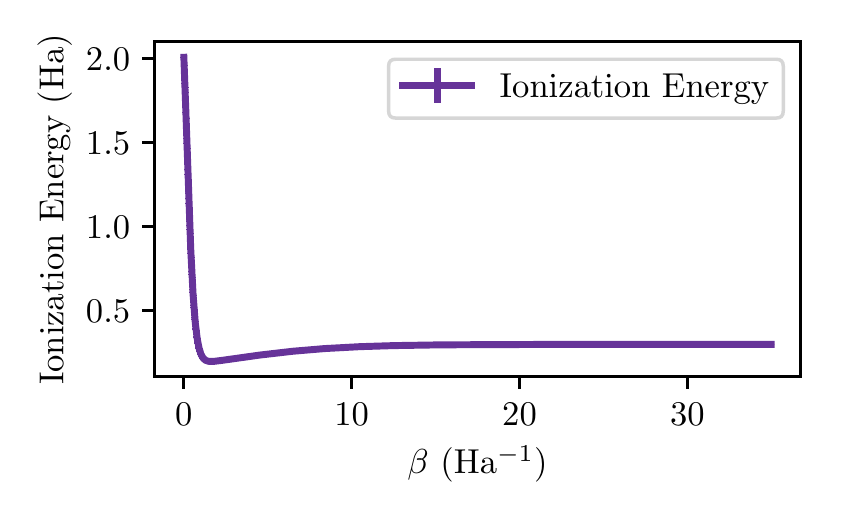}
    
    \caption{The ionization potential for \ce{Be} as a function of inverse temperature ($0.01 \leq \beta \leq 35.00$). Two calculations were performed, one on $\ce{Be}$ and one on $\ce{Be^+}$, both in the MIDI basis set. The propagated error is plotted, but is invisible on this scale. }
    \label{be_ion}
\end{figure}

In \reffig{be_ion} we plot the energy difference, $\Delta U = U_{\ce{Be+}}- U_{\ce{Be}}$, representing the first ionization potential of \ce{Be}. As can be seen from the plot, the calculated ionization energy is less than 0.5 Ha for most values of $\beta$, which indicates a method with milliHartree accuracy is needed in order to accurately capture energy differences such as ionization potential We hope to extend this procedure in future work by investigating high oxidation states of transition metals.  

We have shown \iDMQMC is well-equipped for calculating properties than involve energy differences. The stochastic nature of \iDMQMC gives good control of the noise in the calculation, allowing for calculated differences between energy to be physically significant.

\section{Conclusions}

The main achievement of the DMQMC method to date has been been to treat model systems such as the 2d-Heisenberg model\cite{blunt_density-matrix_2014} and the uniform electron gas.\cite{malone_accurate_2016} In the latter, DMQMC was recently used to benchmark the warm dense electron gas with implications to finite density temperature functionals.\cite{malone_accurate_2016}
In these systems it has shown that Hilbert spaces of significant system sizes can be treated with exact-on-average accuracy. 

In this paper, we took a first look at DMQMC as it applies to ab initio systems. 
We aimed to reproduce the same series of calculations presented by Rubenstein et al. who used FT-AFQMC on \ce{Be}, \ce{H2O}, and \ce{H10} at near-equilibrium and stretched geometries. 
We successfully showed that i-DMQMC can reproduce FCI results within 1 millihartree (stochastic and systematic error) for \ce{Be} and \ce{H2O} for a wide range of inverse temperatures. 
The initiator error was shown to be well-controlled by raising the walker population size. 
In comparing i-DMQMC with FT-AFQMC, we were able to benchmark the difference between the grand canonical ensemble and canonical ensemble energies for finite systems, specifically, for \ce{H2O} and \ce{H10} at near-equilibrium and stretched geometries.
Finally, we demonstrated that it was possible to take numerical differences between i-DMQMC results since the stochastic noise is sufficiently well-controlled.

The DMQMC method is a close relation to FCIQMC\cite{booth_fermion_2009} and, as such, it remains an open question as to whether DMQMC can have the same success in treating systems with larger system sizes. 
We note that we have left open the question of how the method scales, although it is anticipated that it will scale in a similar fashion to FCIQMC.\cite{cleland_taming_2012}
In the early days of FCIQMC, it went through significant development that saw it reduce the scaling of the method considerably and it is hoped that with sufficient attention DMQMC may be able to follow suit. 
Another way that DMQMC could find a niche is if it can treat (and benchmark) finite temperature energies for solids, which appears to be one of the large open questions in finite temperature electronic structure theory.
We are indeed interested in this question and it will be investigated in future work which is forthcoming.

\section{Acknowledgements}
JJS, SKR, and HRP acknowledge the University of Iowa for funding.  
The work of FDM was performed under the auspices of the U.S. Department of Energy
(DOE) by LLNL under Contract No. DE-AC52-07NA27344.  Computer
time was provided by the Livermore Computing Facilities and the University of Iowa Informatics Initiative.
The code used throughout this work was HANDE (www.hande.org.uk).
We gratefully acknowledge Maciej Jarocki for preliminary work in this area and Matthew Foulkes for discussions at Imperial College London. We would also like to thank Brenda Rubenstein and Yuan Liu for discussions and access to the 1.786 \AA \ce{H10} and 2.4 \AA \ce{H10} integral files used in their paper.\cite{liu_ab_2018}

\bibliography{hrp_bib,hrp_bib_dmqmc,pyscf,molpro,zo}%

\begin{mcitethebibliography}{66}
\providecommand*\natexlab[1]{#1}
\providecommand*\mciteSetBstSublistMode[1]{}
\providecommand*\mciteSetBstMaxWidthForm[2]{}
\providecommand*\mciteBstWouldAddEndPuncttrue
  {\def\EndOfBibitem{\unskip.}}
\providecommand*\mciteBstWouldAddEndPunctfalse
  {\let\EndOfBibitem\relax}
\providecommand*\mciteSetBstMidEndSepPunct[3]{}
\providecommand*\mciteSetBstSublistLabelBeginEnd[3]{}
\providecommand*\EndOfBibitem{}
\mciteSetBstSublistMode{f}
\mciteSetBstMaxWidthForm{subitem}{(\alph{mcitesubitemcount})}
\mciteSetBstSublistLabelBeginEnd
  {\mcitemaxwidthsubitemform\space}
  {\relax}
  {\relax}

\bibitem[Graziani \latin{et~al.}(2014)Graziani, Desjarlais, Redmer, and
  Trickey]{graziani_frontiers_2014}
Graziani,~F.; Desjarlais,~M.~P.; Redmer,~R.; Trickey,~S.~B. \emph{Frontiers and
  {Challenges} in {Warm} {Dense} {Matter}}; Springer Science \& Business, 2014;
  Google-Books-ID: Hdm4BAAAQBAJ\relax
\mciteBstWouldAddEndPuncttrue
\mciteSetBstMidEndSepPunct{\mcitedefaultmidpunct}
{\mcitedefaultendpunct}{\mcitedefaultseppunct}\relax
\EndOfBibitem
\bibitem[Dornheim \latin{et~al.}(2018)Dornheim, Groth, and
  Bonitz]{dornheim_review}
Dornheim,~T.; Groth,~S.; Bonitz,~M. The uniform electron gas at warm dense
  matter conditions. \emph{Physics Reports} \textbf{2018}, \emph{744}, 1 --
  86\relax
\mciteBstWouldAddEndPuncttrue
\mciteSetBstMidEndSepPunct{\mcitedefaultmidpunct}
{\mcitedefaultendpunct}{\mcitedefaultseppunct}\relax
\EndOfBibitem
\bibitem[Benuzzi-Mounaix \latin{et~al.}(2014)Benuzzi-Mounaix, Mazevet, Ravasio,
  Vinci, Denoeud, Koenig, Amadou, Brambrink, Festa, Levy, Harmand, Brygoo,
  Huser, Recoules, Bouchet, Morard, Guyot, Resseguier, Myanishi, Ozaki,
  Dorchies, Gaudin, Leguay, Peyrusse, Henry, Raffestin, Pape, Smith, and
  Musella]{benuzzi-mounaix_progress_2014}
Benuzzi-Mounaix,~A.; Mazevet,~S.; Ravasio,~A.; Vinci,~T.; Denoeud,~A.;
  Koenig,~M.; Amadou,~N.; Brambrink,~E.; Festa,~F.; Levy,~A.; Harmand,~M.;
  Brygoo,~S.; Huser,~G.; Recoules,~V.; Bouchet,~J.; Morard,~G.; Guyot,~F.;
  Resseguier,~T.~d.; Myanishi,~K.; Ozaki,~N.; Dorchies,~F.; Gaudin,~J.;
  Leguay,~P.~M.; Peyrusse,~O.; Henry,~O.; Raffestin,~D.; Pape,~S.~L.;
  Smith,~R.; Musella,~R. Progress in warm dense matter study with applications
  to planetology. \emph{Physica Scripta} \textbf{2014}, \emph{T161}, 014060,
  bibtex: benuzzi-mounaix\_progress\_2014\relax
\mciteBstWouldAddEndPuncttrue
\mciteSetBstMidEndSepPunct{\mcitedefaultmidpunct}
{\mcitedefaultendpunct}{\mcitedefaultseppunct}\relax
\EndOfBibitem
\bibitem[Koenig \latin{et~al.}(2005)Koenig, Benuzzi-Mounaix, Ravasio, Vinci,
  Ozaki, Lepape, Batani, Huser, Hall, Hicks, MacKinnon, Patel, Park, Boehly,
  Borghesi, Kar, and Romagnani]{koenig_progress_2005}
Koenig,~M.; Benuzzi-Mounaix,~A.; Ravasio,~A.; Vinci,~T.; Ozaki,~N.; Lepape,~S.;
  Batani,~D.; Huser,~G.; Hall,~T.; Hicks,~D.; MacKinnon,~A.; Patel,~P.;
  Park,~H.~S.; Boehly,~T.; Borghesi,~M.; Kar,~S.; Romagnani,~L. Progress in the
  study of warm dense matter. \emph{Plasma Physics and Controlled Fusion}
  \textbf{2005}, \emph{47}, B441\relax
\mciteBstWouldAddEndPuncttrue
\mciteSetBstMidEndSepPunct{\mcitedefaultmidpunct}
{\mcitedefaultendpunct}{\mcitedefaultseppunct}\relax
\EndOfBibitem
\bibitem[Mukherjee \latin{et~al.}(2013)Mukherjee, Libisch, Large, Neumann,
  Brown, Cheng, Lassiter, Carter, Nordlander, and Halas]{mukherjee_hot_2013-1}
Mukherjee,~S.; Libisch,~F.; Large,~N.; Neumann,~O.; Brown,~L.~V.; Cheng,~J.;
  Lassiter,~J.~B.; Carter,~E.~A.; Nordlander,~P.; Halas,~N.~J. Hot {Electrons}
  {Do} the {Impossible}: {Plasmon}-{Induced} {Dissociation} of {H}
  $_{\textrm{2}}$ on {Au}. \emph{Nano Letters} \textbf{2013}, \emph{13},
  240--247\relax
\mciteBstWouldAddEndPuncttrue
\mciteSetBstMidEndSepPunct{\mcitedefaultmidpunct}
{\mcitedefaultendpunct}{\mcitedefaultseppunct}\relax
\EndOfBibitem
\bibitem[Ernstorfer \latin{et~al.}(2009)Ernstorfer, Harb, Hebeisen, Sciaini,
  Dartigalongue, and Miller]{ernstorfer_formation_2009}
Ernstorfer,~R.; Harb,~M.; Hebeisen,~C.~T.; Sciaini,~G.; Dartigalongue,~T.;
  Miller,~R. J.~D. The {Formation} of {Warm} {Dense} {Matter}: {Experimental}
  {Evidence} for {Electronic} {Bond} {Hardening} in {Gold}. \emph{Science}
  \textbf{2009}, \emph{323}, 1033--1037, bibtex:
  ernstorfer\_formation\_2009\relax
\mciteBstWouldAddEndPuncttrue
\mciteSetBstMidEndSepPunct{\mcitedefaultmidpunct}
{\mcitedefaultendpunct}{\mcitedefaultseppunct}\relax
\EndOfBibitem
\bibitem[Fletcher \latin{et~al.}(2015)Fletcher, Lee, D{\"o}ppner, Galtier,
  Nagler, Heimann, Fortmann, LePape, Ma, Millot, Pak, Turnbull, Chapman,
  Gericke, Vorberger, White, Gregori, Wei, Barbrel, Falcone, Kao, Nuhn, Welch,
  Zastrau, Neumayer, Hastings, and Glenzer]{glenzer_slac_15}
Fletcher,~L.~B.; Lee,~H.~J.; D{\"o}ppner,~T.; Galtier,~E.; Nagler,~B.;
  Heimann,~P.; Fortmann,~C.; LePape,~S.; Ma,~T.; Millot,~M.; Pak,~A.;
  Turnbull,~D.; Chapman,~D.~A.; Gericke,~D.~O.; Vorberger,~J.; White,~T.;
  Gregori,~G.; Wei,~M.; Barbrel,~B.; Falcone,~R.~W.; Kao,~C.~C.; Nuhn,~H.;
  Welch,~J.; Zastrau,~U.; Neumayer,~P.; Hastings,~J.~B.; Glenzer,~S.~H.
  Ultrabright X-ray laser scattering for dynamic warm dense matter physics.
  \emph{Nat. Photonics} \textbf{2015}, \emph{9}, 274\relax
\mciteBstWouldAddEndPuncttrue
\mciteSetBstMidEndSepPunct{\mcitedefaultmidpunct}
{\mcitedefaultendpunct}{\mcitedefaultseppunct}\relax
\EndOfBibitem
\bibitem[Mermin(1965)]{mermin_tdft}
Mermin,~N.~D. Thermal Properties of the Inhomogeneous Electron Gas. \emph{Phys.
  Rev.} \textbf{1965}, \emph{137}, A1441--A1443\relax
\mciteBstWouldAddEndPuncttrue
\mciteSetBstMidEndSepPunct{\mcitedefaultmidpunct}
{\mcitedefaultendpunct}{\mcitedefaultseppunct}\relax
\EndOfBibitem
\bibitem[Zhang \latin{et~al.}(2017)Zhang, Driver, Soubiran, and
  Militzer]{zhang_hydrocarbons_17}
Zhang,~S.; Driver,~K.~P.; Soubiran,~F.; Militzer,~B. First-principles equation
  of state and shock compression predictions of warm dense hydrocarbons.
  \emph{Phys. Rev. E} \textbf{2017}, \emph{96}, 013204\relax
\mciteBstWouldAddEndPuncttrue
\mciteSetBstMidEndSepPunct{\mcitedefaultmidpunct}
{\mcitedefaultendpunct}{\mcitedefaultseppunct}\relax
\EndOfBibitem
\bibitem[Karasiev \latin{et~al.}(2016)Karasiev, Calder\'{\i}n, and
  Trickey]{karasiev_importance_2016}
Karasiev,~V.~V.; Calder\'{\i}n,~L.; Trickey,~S.~B. Importance of
  finite-temperature exchange correlation for warm dense matter calculations.
  \emph{Phys. Rev. E} \textbf{2016}, \emph{93}, 063207\relax
\mciteBstWouldAddEndPuncttrue
\mciteSetBstMidEndSepPunct{\mcitedefaultmidpunct}
{\mcitedefaultendpunct}{\mcitedefaultseppunct}\relax
\EndOfBibitem
\bibitem[Karasiev \latin{et~al.}(2014)Karasiev, Sjostrom, Dufty, and
  Trickey]{karasiev_tlda_14}
Karasiev,~V.~V.; Sjostrom,~T.; Dufty,~J.; Trickey,~S.~B. Accurate Homogeneous
  Electron Gas Exchange-Correlation Free Energy for Local Spin-Density
  Calculations. \emph{Phys. Rev. Lett.} \textbf{2014}, \emph{112}, 076403\relax
\mciteBstWouldAddEndPuncttrue
\mciteSetBstMidEndSepPunct{\mcitedefaultmidpunct}
{\mcitedefaultendpunct}{\mcitedefaultseppunct}\relax
\EndOfBibitem
\bibitem[Groth \latin{et~al.}(2017)Groth, Dornheim, Sjostrom, Malone, Foulkes,
  and Bonitz]{groth_fxc_17}
Groth,~S.; Dornheim,~T.; Sjostrom,~T.; Malone,~F.~D.; Foulkes,~W. M.~C.;
  Bonitz,~M. Ab initio Exchange-Correlation Free Energy of the Uniform Electron
  Gas at Warm Dense Matter Conditions. \emph{Phys. Rev. Lett.} \textbf{2017},
  \emph{119}, 135001\relax
\mciteBstWouldAddEndPuncttrue
\mciteSetBstMidEndSepPunct{\mcitedefaultmidpunct}
{\mcitedefaultendpunct}{\mcitedefaultseppunct}\relax
\EndOfBibitem
\bibitem[Karasiev \latin{et~al.}(2019)Karasiev, Hu, Zaghoo, and
  Boehly]{karasiev_xc_deuterium}
Karasiev,~V.~V.; Hu,~S.~X.; Zaghoo,~M.; Boehly,~T.~R. Exchange-correlation
  thermal effects in shocked deuterium: Softening the principal Hugoniot and
  thermophysical properties. \emph{Phys. Rev. B} \textbf{2019}, \emph{99},
  214110\relax
\mciteBstWouldAddEndPuncttrue
\mciteSetBstMidEndSepPunct{\mcitedefaultmidpunct}
{\mcitedefaultendpunct}{\mcitedefaultseppunct}\relax
\EndOfBibitem
\bibitem[Mermin(1963)]{mermin_thf}
Mermin,~N.~D. {Stability of the thermal Hartree-Fock approximation}. \emph{Ann.
  Phys.} \textbf{1963}, \emph{21}, 99--121\relax
\mciteBstWouldAddEndPuncttrue
\mciteSetBstMidEndSepPunct{\mcitedefaultmidpunct}
{\mcitedefaultendpunct}{\mcitedefaultseppunct}\relax
\EndOfBibitem
\bibitem[Kou and Hirata(2014)Kou, and Hirata]{kou_finite-temperature_2014}
Kou,~Z.; Hirata,~S. Finite-temperature full configuration interaction.
  \emph{Theoretical Chemistry Accounts} \textbf{2014}, \emph{133}\relax
\mciteBstWouldAddEndPuncttrue
\mciteSetBstMidEndSepPunct{\mcitedefaultmidpunct}
{\mcitedefaultendpunct}{\mcitedefaultseppunct}\relax
\EndOfBibitem
\bibitem[He \latin{et~al.}(2014)He, Ryu, and
  Hirata]{he_finite-temperature_2014}
He,~X.; Ryu,~S.; Hirata,~S. Finite-temperature second-order many-body
  perturbation and {Hartree}–{Fock} theories for one-dimensional solids: {An}
  application to {Peierls} and charge-density-wave transitions in conjugated
  polymers. \emph{The Journal of Chemical Physics} \textbf{2014}, \emph{140},
  024702\relax
\mciteBstWouldAddEndPuncttrue
\mciteSetBstMidEndSepPunct{\mcitedefaultmidpunct}
{\mcitedefaultendpunct}{\mcitedefaultseppunct}\relax
\EndOfBibitem
\bibitem[Hermes and Hirata(2015)Hermes, and
  Hirata]{hermes_finite-temperature_2015}
Hermes,~M.~R.; Hirata,~S. Finite-temperature coupled-cluster, many-body
  perturbation, and restricted and unrestricted {Hartree}–{Fock} study on
  one-dimensional solids: {Luttinger} liquids, {Peierls} transitions, and spin-
  and charge-density waves. \emph{The Journal of Chemical Physics}
  \textbf{2015}, \emph{143}, 102818\relax
\mciteBstWouldAddEndPuncttrue
\mciteSetBstMidEndSepPunct{\mcitedefaultmidpunct}
{\mcitedefaultendpunct}{\mcitedefaultseppunct}\relax
\EndOfBibitem
\bibitem[Rusakov and Zgid(2016)Rusakov, and Zgid]{rusakov_self-consistent_2016}
Rusakov,~A.~A.; Zgid,~D. Self-consistent second-order {Green}’s function
  perturbation theory for periodic systems. \emph{The Journal of Chemical
  Physics} \textbf{2016}, \emph{144}, 054106\relax
\mciteBstWouldAddEndPuncttrue
\mciteSetBstMidEndSepPunct{\mcitedefaultmidpunct}
{\mcitedefaultendpunct}{\mcitedefaultseppunct}\relax
\EndOfBibitem
\bibitem[Hummel(2018)]{hummel_finite_2018}
Hummel,~F. Finite {Temperature} {Coupled} {Cluster} {Theories} for {Extended}
  {Systems}. \emph{Journal of Chemical Theory and Computation} \textbf{2018},
  \emph{14}, 6505--6514\relax
\mciteBstWouldAddEndPuncttrue
\mciteSetBstMidEndSepPunct{\mcitedefaultmidpunct}
{\mcitedefaultendpunct}{\mcitedefaultseppunct}\relax
\EndOfBibitem
\bibitem[White and Chan(2018)White, and Chan]{white_time-dependent_2018}
White,~A.~F.; Chan,~G. K.-L. A {Time}-{Dependent} {Formulation} of
  {Coupled}-{Cluster} {Theory} for {Many}-{Fermion} {Systems} at {Finite}
  {Temperature}. \emph{Journal of Chemical Theory and Computation}
  \textbf{2018}, \emph{14}, 5690--5700\relax
\mciteBstWouldAddEndPuncttrue
\mciteSetBstMidEndSepPunct{\mcitedefaultmidpunct}
{\mcitedefaultendpunct}{\mcitedefaultseppunct}\relax
\EndOfBibitem
\bibitem[Harsha \latin{et~al.}(2019)Harsha, Henderson, and
  Scuseria]{harsha_thermofield_2019}
Harsha,~G.; Henderson,~T.~M.; Scuseria,~G.~E. Thermofield theory for
  finite-temperature quantum chemistry. \emph{The Journal of Chemical Physics}
  \textbf{2019}, \emph{150}, 154109\relax
\mciteBstWouldAddEndPuncttrue
\mciteSetBstMidEndSepPunct{\mcitedefaultmidpunct}
{\mcitedefaultendpunct}{\mcitedefaultseppunct}\relax
\EndOfBibitem
\bibitem[Harsha \latin{et~al.}(2019)Harsha, Henderson, and
  Scuseria]{harsha_thermofield_2019-1}
Harsha,~G.; Henderson,~T.~M.; Scuseria,~G.~E. Thermofield theory for
  finite-temperature coupled cluster. \emph{Journal of Chemical Theory and
  Computation} \textbf{2019}, \relax
\mciteBstWouldAddEndPunctfalse
\mciteSetBstMidEndSepPunct{\mcitedefaultmidpunct}
{}{\mcitedefaultseppunct}\relax
\EndOfBibitem
\bibitem[Sanyal \latin{et~al.}(1992)Sanyal, Mandal, and
  Mukherjee]{sanyal_thermal_1992}
Sanyal,~G.; Mandal,~S.~H.; Mukherjee,~D. Thermal averaging in quantum many-body
  systems: a non-perturbative thermal cluster cumulant approach. \emph{Chemical
  Physics Letters} \textbf{1992}, \emph{192}, 55--61\relax
\mciteBstWouldAddEndPuncttrue
\mciteSetBstMidEndSepPunct{\mcitedefaultmidpunct}
{\mcitedefaultendpunct}{\mcitedefaultseppunct}\relax
\EndOfBibitem
\bibitem[Ceperley(1995)]{ceperley_pimc}
Ceperley,~D.~M. Path integrals in the theory of condensed helium. \emph{Rev.
  Mod. Phys.} \textbf{1995}, \emph{67}, 279--355\relax
\mciteBstWouldAddEndPuncttrue
\mciteSetBstMidEndSepPunct{\mcitedefaultmidpunct}
{\mcitedefaultendpunct}{\mcitedefaultseppunct}\relax
\EndOfBibitem
\bibitem[Driver and Militzer(2012)Driver, and
  Militzer]{driver_all-electron_2012}
Driver,~K.~P.; Militzer,~B. All-{Electron} {Path} {Integral} {Monte} {Carlo}
  {Simulations} of {Warm} {Dense} {Matter}: {Application} to {Water} and
  {Carbon} {Plasmas}. \emph{Phys. Rev. Lett.} \textbf{2012}, \emph{108},
  115502\relax
\mciteBstWouldAddEndPuncttrue
\mciteSetBstMidEndSepPunct{\mcitedefaultmidpunct}
{\mcitedefaultendpunct}{\mcitedefaultseppunct}\relax
\EndOfBibitem
\bibitem[Driver \latin{et~al.}(2015)Driver, Soubiran, Zhang, and
  Militzer]{driver_wdm_oxygen}
Driver,~K.~P.; Soubiran,~F.; Zhang,~S.; Militzer,~B. First-principles equation
  of state and electronic properties of warm dense oxygen. \emph{The Journal of
  Chemical Physics} \textbf{2015}, \emph{143}, 164507\relax
\mciteBstWouldAddEndPuncttrue
\mciteSetBstMidEndSepPunct{\mcitedefaultmidpunct}
{\mcitedefaultendpunct}{\mcitedefaultseppunct}\relax
\EndOfBibitem
\bibitem[Ceperley(1991)]{ceperley_fermion_nodes}
Ceperley,~D.~M. Fermion nodes. \emph{J. Stat. Phys.} \textbf{1991}, \emph{63},
  1237--1267\relax
\mciteBstWouldAddEndPuncttrue
\mciteSetBstMidEndSepPunct{\mcitedefaultmidpunct}
{\mcitedefaultendpunct}{\mcitedefaultseppunct}\relax
\EndOfBibitem
\bibitem[Foulkes \latin{et~al.}(2001)Foulkes, Mitas, Needs, and
  Rajagopal]{foulkes_quantum_2001}
Foulkes,~W. M.~C.; Mitas,~L.; Needs,~R.~J.; Rajagopal,~G. Quantum {Monte}
  {Carlo} simulations of solids. \emph{Reviews of Modern Physics}
  \textbf{2001}, \emph{73}, 33--83\relax
\mciteBstWouldAddEndPuncttrue
\mciteSetBstMidEndSepPunct{\mcitedefaultmidpunct}
{\mcitedefaultendpunct}{\mcitedefaultseppunct}\relax
\EndOfBibitem
\bibitem[Brown \latin{et~al.}(2013)Brown, Clark, DuBois, and
  Ceperley]{brown_path-integral_2013}
Brown,~E.~W.; Clark,~B.~K.; DuBois,~J.~L.; Ceperley,~D.~M. Path-{Integral}
  {Monte} {Carlo} {Simulation} of the {Warm} {Dense} {Homogeneous} {Electron}
  {Gas}. \emph{Phys. Rev. Lett.} \textbf{2013}, \emph{110}, 146405\relax
\mciteBstWouldAddEndPuncttrue
\mciteSetBstMidEndSepPunct{\mcitedefaultmidpunct}
{\mcitedefaultendpunct}{\mcitedefaultseppunct}\relax
\EndOfBibitem
\bibitem[Schoof \latin{et~al.}(2015)Schoof, Groth, Vorberger, and
  Bonitz]{schoof_textitab_2015}
Schoof,~T.; Groth,~S.; Vorberger,~J.; Bonitz,~M. {\textbackslash}textit\{{Ab}
  {Initio}\} {Thermodynamic} {Results} for the {Degenerate} {Electron} {Gas} at
  {Finite} {Temperature}. \emph{Phys. Rev. Lett.} \textbf{2015}, \emph{115},
  130402\relax
\mciteBstWouldAddEndPuncttrue
\mciteSetBstMidEndSepPunct{\mcitedefaultmidpunct}
{\mcitedefaultendpunct}{\mcitedefaultseppunct}\relax
\EndOfBibitem
\bibitem[Malone \latin{et~al.}(2016)Malone, Blunt, Brown, Lee, Spencer,
  Foulkes, and Shepherd]{malone_accurate_2016}
Malone,~F.~D.; Blunt,~N.; Brown,~E.~W.; Lee,~D.; Spencer,~J.; Foulkes,~W.;
  Shepherd,~J.~J. Accurate {Exchange}-{Correlation} {Energies} for the {Warm}
  {Dense} {Electron} {Gas}. \emph{Physical Review Letters} \textbf{2016},
  \emph{117}\relax
\mciteBstWouldAddEndPuncttrue
\mciteSetBstMidEndSepPunct{\mcitedefaultmidpunct}
{\mcitedefaultendpunct}{\mcitedefaultseppunct}\relax
\EndOfBibitem
\bibitem[Dornheim \latin{et~al.}(2015)Dornheim, Groth, Filinov, and
  Bonitz]{dornheim_pbpimc}
Dornheim,~T.; Groth,~S.; Filinov,~A.; Bonitz,~M. {Permutation blocking path
  integral Monte Carlo: a highly efficient approach to the simulation of
  strongly degenerate non-ideal fermions}. \emph{New J. of Phys.}
  \textbf{2015}, \emph{17}, 73017\relax
\mciteBstWouldAddEndPuncttrue
\mciteSetBstMidEndSepPunct{\mcitedefaultmidpunct}
{\mcitedefaultendpunct}{\mcitedefaultseppunct}\relax
\EndOfBibitem
\bibitem[DuBois \latin{et~al.}()DuBois, Brown, and Alder]{dubois_sign}
DuBois,~J.~L.; Brown,~E.~W.; Alder,~B.~J. \emph{Advances in the Computational
  Sciences}; Chapter Chapter 13, pp 184--192\relax
\mciteBstWouldAddEndPuncttrue
\mciteSetBstMidEndSepPunct{\mcitedefaultmidpunct}
{\mcitedefaultendpunct}{\mcitedefaultseppunct}\relax
\EndOfBibitem
\bibitem[Schoof \latin{et~al.}(2011)Schoof, Bonitz, Filinov, Hochstuhl, and
  Dufty]{schoof_cpimc}
Schoof,~T.; Bonitz,~M.; Filinov,~A.; Hochstuhl,~D.; Dufty,~J.~W. {Configuration
  Path Integral Monte Carlo}. \emph{Contr. Plasma Phys.} \textbf{2011},
  \emph{51}, 687--697\relax
\mciteBstWouldAddEndPuncttrue
\mciteSetBstMidEndSepPunct{\mcitedefaultmidpunct}
{\mcitedefaultendpunct}{\mcitedefaultseppunct}\relax
\EndOfBibitem
\bibitem[Blankenbecler \latin{et~al.}(1981)Blankenbecler, Scalapino, and
  Sugar]{blankenbecler_dqmc_1}
Blankenbecler,~R.; Scalapino,~D.~J.; Sugar,~R.~L. Monte Carlo calculations of
  coupled boson-fermion systems. I. \emph{Phys. Rev. D} \textbf{1981},
  \emph{24}, 2278--2286\relax
\mciteBstWouldAddEndPuncttrue
\mciteSetBstMidEndSepPunct{\mcitedefaultmidpunct}
{\mcitedefaultendpunct}{\mcitedefaultseppunct}\relax
\EndOfBibitem
\bibitem[Scalapino and Sugar(1981)Scalapino, and Sugar]{scalapino_dqmc}
Scalapino,~D.~J.; Sugar,~R.~L. Method for Performing Monte Carlo Calculations
  for Systems with Fermions. \emph{Phys. Rev. Lett.} \textbf{1981}, \emph{46},
  519--521\relax
\mciteBstWouldAddEndPuncttrue
\mciteSetBstMidEndSepPunct{\mcitedefaultmidpunct}
{\mcitedefaultendpunct}{\mcitedefaultseppunct}\relax
\EndOfBibitem
\bibitem[Zhang(1999)]{zhang_ftafqmc_99}
Zhang,~S. Finite-Temperature Monte Carlo Calculations for Systems with
  Fermions. \emph{Phys. Rev. Lett.} \textbf{1999}, \emph{83}, 2777--2780\relax
\mciteBstWouldAddEndPuncttrue
\mciteSetBstMidEndSepPunct{\mcitedefaultmidpunct}
{\mcitedefaultendpunct}{\mcitedefaultseppunct}\relax
\EndOfBibitem
\bibitem[Rubenstein \latin{et~al.}(2012)Rubenstein, Zhang, and
  Reichman]{rubenstein_finite-temperature_2012}
Rubenstein,~B.~M.; Zhang,~S.; Reichman,~D.~R. Finite-temperature
  auxiliary-field quantum {Monte} {Carlo} technique for {Bose}-{Fermi}
  mixtures. \emph{Phys. Rev. A} \textbf{2012}, \emph{86}, 053606\relax
\mciteBstWouldAddEndPuncttrue
\mciteSetBstMidEndSepPunct{\mcitedefaultmidpunct}
{\mcitedefaultendpunct}{\mcitedefaultseppunct}\relax
\EndOfBibitem
\bibitem[LeBlanc \latin{et~al.}(2015)LeBlanc, Antipov, Becca, Bulik, Chan,
  Chung, Deng, Ferrero, Henderson, Jim\'enez-Hoyos, Kozik, Liu, Millis,
  Prokof'ev, Qin, Scuseria, Shi, Svistunov, Tocchio, Tupitsyn, White, Zhang,
  Zheng, Zhu, and Gull]{simons_hubbard_2d}
LeBlanc,~J. P.~F.; Antipov,~A.~E.; Becca,~F.; Bulik,~I.~W.; Chan,~G. K.-L.;
  Chung,~C.-M.; Deng,~Y.; Ferrero,~M.; Henderson,~T.~M.;
  Jim\'enez-Hoyos,~C.~A.; Kozik,~E.; Liu,~X.-W.; Millis,~A.~J.;
  Prokof'ev,~N.~V.; Qin,~M.; Scuseria,~G.~E.; Shi,~H.; Svistunov,~B.~V.;
  Tocchio,~L.~F.; Tupitsyn,~I.~S.; White,~S.~R.; Zhang,~S.; Zheng,~B.-X.;
  Zhu,~Z.; Gull,~E. Solutions of the Two-Dimensional Hubbard Model: Benchmarks
  and Results from a Wide Range of Numerical Algorithms. \emph{Phys. Rev. X}
  \textbf{2015}, \emph{5}, 041041\relax
\mciteBstWouldAddEndPuncttrue
\mciteSetBstMidEndSepPunct{\mcitedefaultmidpunct}
{\mcitedefaultendpunct}{\mcitedefaultseppunct}\relax
\EndOfBibitem
\bibitem[Motta \latin{et~al.}(2017)Motta, Ceperley, Chan, Gomez, Gull, Guo,
  Jiménez-Hoyos, Lan, Li, Ma, Millis, Prokof’ev, Ray, Scuseria, Sorella,
  Stoudenmire, Sun, Tupitsyn, White, Zgid, Zhang, and {Simons Collaboration on
  the Many-Electron Problem}]{motta_towards_2017}
Motta,~M.; Ceperley,~D.~M.; Chan,~G. K.-L.; Gomez,~J.~A.; Gull,~E.; Guo,~S.;
  Jiménez-Hoyos,~C.~A.; Lan,~T.~N.; Li,~J.; Ma,~F.; Millis,~A.~J.;
  Prokof’ev,~N.~V.; Ray,~U.; Scuseria,~G.~E.; Sorella,~S.;
  Stoudenmire,~E.~M.; Sun,~Q.; Tupitsyn,~I.~S.; White,~S.~R.; Zgid,~D.;
  Zhang,~S.; {Simons Collaboration on the Many-Electron Problem}, Towards the
  {Solution} of the {Many}-{Electron} {Problem} in {Real} {Materials}:
  {Equation} of {State} of the {Hydrogen} {Chain} with {State}-of-the-{Art}
  {Many}-{Body} {Methods}. \emph{Physical Review X} \textbf{2017},
  \emph{7}\relax
\mciteBstWouldAddEndPuncttrue
\mciteSetBstMidEndSepPunct{\mcitedefaultmidpunct}
{\mcitedefaultendpunct}{\mcitedefaultseppunct}\relax
\EndOfBibitem
\bibitem[Motta and Zhang(2018)Motta, and Zhang]{motta_review}
Motta,~M.; Zhang,~S. Ab initio computations of molecular systems by the
  auxiliary-field quantum Monte Carlo method. \emph{WIREs Comput. Mol. Sci.}
  \textbf{2018}, \emph{8}, e1364\relax
\mciteBstWouldAddEndPuncttrue
\mciteSetBstMidEndSepPunct{\mcitedefaultmidpunct}
{\mcitedefaultendpunct}{\mcitedefaultseppunct}\relax
\EndOfBibitem
\bibitem[Zhang and Krakauer(2003)Zhang, and Krakauer]{zhang_phaseless}
Zhang,~S.; Krakauer,~H. Quantum Monte Carlo Method using Phase-Free Random
  Walks with Slater Determinants. \emph{Phys. Rev. Lett.} \textbf{2003},
  \emph{90}, 136401\relax
\mciteBstWouldAddEndPuncttrue
\mciteSetBstMidEndSepPunct{\mcitedefaultmidpunct}
{\mcitedefaultendpunct}{\mcitedefaultseppunct}\relax
\EndOfBibitem
\bibitem[Liu \latin{et~al.}(2018)Liu, Cho, and Rubenstein]{liu_ab_2018}
Liu,~Y.; Cho,~M.; Rubenstein,~B. \textit{{Ab} {Initio}} {Finite} {Temperature}
  {Auxiliary} {Field} {Quantum} {Monte} {Carlo}. \emph{Journal of Chemical
  Theory and Computation} \textbf{2018}, \emph{14}, 4722--4732\relax
\mciteBstWouldAddEndPuncttrue
\mciteSetBstMidEndSepPunct{\mcitedefaultmidpunct}
{\mcitedefaultendpunct}{\mcitedefaultseppunct}\relax
\EndOfBibitem
\bibitem[Blunt \latin{et~al.}(2014)Blunt, Rogers, Spencer, and
  Foulkes]{blunt_density-matrix_2014}
Blunt,~N.~S.; Rogers,~T.~W.; Spencer,~J.~S.; Foulkes,~W. M.~C. Density-matrix
  quantum {Monte} {Carlo} method. \emph{Physical Review B} \textbf{2014},
  \emph{89}\relax
\mciteBstWouldAddEndPuncttrue
\mciteSetBstMidEndSepPunct{\mcitedefaultmidpunct}
{\mcitedefaultendpunct}{\mcitedefaultseppunct}\relax
\EndOfBibitem
\bibitem[Booth \latin{et~al.}(2009)Booth, Thom, and Alavi]{booth_fermion_2009}
Booth,~G.~H.; Thom,~A. J.~W.; Alavi,~A. Fermion {Monte} {Carlo} without fixed
  nodes: {A} game of life, death, and annihilation in {Slater} determinant
  space. \emph{The Journal of Chemical Physics} \textbf{2009}, \emph{131},
  054106\relax
\mciteBstWouldAddEndPuncttrue
\mciteSetBstMidEndSepPunct{\mcitedefaultmidpunct}
{\mcitedefaultendpunct}{\mcitedefaultseppunct}\relax
\EndOfBibitem
\bibitem[Malone \latin{et~al.}(2015)Malone, Blunt, Shepherd, Lee, Spencer, and
  Foulkes]{malone_interaction_2015}
Malone,~F.~D.; Blunt,~N.~S.; Shepherd,~J.~J.; Lee,~D. K.~K.; Spencer,~J.~S.;
  Foulkes,~W. M.~C. Interaction picture density matrix quantum {Monte} {Carlo}.
  \emph{The Journal of Chemical Physics} \textbf{2015}, \emph{143},
  044116\relax
\mciteBstWouldAddEndPuncttrue
\mciteSetBstMidEndSepPunct{\mcitedefaultmidpunct}
{\mcitedefaultendpunct}{\mcitedefaultseppunct}\relax
\EndOfBibitem
\bibitem[Cleland \latin{et~al.}(2010)Cleland, Booth, and
  Alavi]{cleland_communications:_2010}
Cleland,~D.; Booth,~G.~H.; Alavi,~A. Communications: {Survival} of the fittest:
  {Accelerating} convergence in full configuration-interaction quantum {Monte}
  {Carlo}. \emph{The Journal of Chemical Physics} \textbf{2010}, \emph{132},
  041103\relax
\mciteBstWouldAddEndPuncttrue
\mciteSetBstMidEndSepPunct{\mcitedefaultmidpunct}
{\mcitedefaultendpunct}{\mcitedefaultseppunct}\relax
\EndOfBibitem
\bibitem[Petruzielo \latin{et~al.}(2012)Petruzielo, Holmes, Changlani,
  Nightingale, and Umrigar]{petruzielo_semistochastic_2012}
Petruzielo,~F.~R.; Holmes,~A.~A.; Changlani,~H.~J.; Nightingale,~M.~P.;
  Umrigar,~C.~J. Semistochastic {Projector} {Monte} {Carlo} {Method}.
  \emph{Physical Review Letters} \textbf{2012}, \emph{109}\relax
\mciteBstWouldAddEndPuncttrue
\mciteSetBstMidEndSepPunct{\mcitedefaultmidpunct}
{\mcitedefaultendpunct}{\mcitedefaultseppunct}\relax
\EndOfBibitem
\bibitem[Blunt \latin{et~al.}(2015)Blunt, Smart, Kersten, Spencer, Booth, and
  Alavi]{blunt_semi-stochastic_2015}
Blunt,~N.~S.; Smart,~S.~D.; Kersten,~J. A.~F.; Spencer,~J.~S.; Booth,~G.~H.;
  Alavi,~A. Semi-stochastic full configuration interaction quantum {Monte}
  {Carlo}: {Developments} and application. \emph{The Journal of Chemical
  Physics} \textbf{2015}, \emph{142}, 184107\relax
\mciteBstWouldAddEndPuncttrue
\mciteSetBstMidEndSepPunct{\mcitedefaultmidpunct}
{\mcitedefaultendpunct}{\mcitedefaultseppunct}\relax
\EndOfBibitem
\bibitem[Holmes \latin{et~al.}(2016)Holmes, Changlani, and
  Umrigar]{holmes_efficient_2016}
Holmes,~A.~A.; Changlani,~H.~J.; Umrigar,~C.~J. Efficient {Heat}-{Bath}
  {Sampling} in {Fock} {Space}. \emph{Journal of Chemical Theory and
  Computation} \textbf{2016}, \emph{12}, 1561--1571\relax
\mciteBstWouldAddEndPuncttrue
\mciteSetBstMidEndSepPunct{\mcitedefaultmidpunct}
{\mcitedefaultendpunct}{\mcitedefaultseppunct}\relax
\EndOfBibitem
\bibitem[Li \latin{et~al.}(2018)Li, Otten, Holmes, Sharma, and
  Umrigar]{li_fast_2018}
Li,~J.; Otten,~M.; Holmes,~A.~A.; Sharma,~S.; Umrigar,~C.~J. Fast
  semistochastic heat-bath configuration interaction. \emph{The Journal of
  Chemical Physics} \textbf{2018}, \emph{149}, 214110\relax
\mciteBstWouldAddEndPuncttrue
\mciteSetBstMidEndSepPunct{\mcitedefaultmidpunct}
{\mcitedefaultendpunct}{\mcitedefaultseppunct}\relax
\EndOfBibitem
\bibitem[Blunt(2018)]{blunt_communication:_2018}
Blunt,~N.~S. Communication: {An} efficient and accurate perturbative correction
  to initiator full configuration interaction quantum {Monte} {Carlo}.
  \emph{The Journal of Chemical Physics} \textbf{2018}, \emph{148},
  221101\relax
\mciteBstWouldAddEndPuncttrue
\mciteSetBstMidEndSepPunct{\mcitedefaultmidpunct}
{\mcitedefaultendpunct}{\mcitedefaultseppunct}\relax
\EndOfBibitem
\bibitem[Blunt \latin{et~al.}(2019)Blunt, Thom, and
  Scott]{blunt_preconditioning_2019}
Blunt,~N.~S.; Thom,~A. J.~W.; Scott,~C. J.~C. Preconditioning and
  {Perturbative} {Estimators} in {Full} {Configuration} {Interaction} {Quantum}
  {Monte} {Carlo}. \emph{Journal of Chemical Theory and Computation}
  \textbf{2019}, \emph{15}, 3537--3551\relax
\mciteBstWouldAddEndPuncttrue
\mciteSetBstMidEndSepPunct{\mcitedefaultmidpunct}
{\mcitedefaultendpunct}{\mcitedefaultseppunct}\relax
\EndOfBibitem
\bibitem[Deustua \latin{et~al.}(2018)Deustua, Magoulas, Shen, and
  Piecuch]{deustua_communication:_2018}
Deustua,~J.~E.; Magoulas,~I.; Shen,~J.; Piecuch,~P. Communication:
  {Approaching} exact quantum chemistry by cluster analysis of full
  configuration interaction quantum {Monte} {Carlo} wave functions. \emph{The
  Journal of Chemical Physics} \textbf{2018}, \emph{149}, 151101\relax
\mciteBstWouldAddEndPuncttrue
\mciteSetBstMidEndSepPunct{\mcitedefaultmidpunct}
{\mcitedefaultendpunct}{\mcitedefaultseppunct}\relax
\EndOfBibitem
\bibitem[Deustua \latin{et~al.}(2019)Deustua, Yuwono, Shen, and
  Piecuch]{deustua_accurate_2019}
Deustua,~J.~E.; Yuwono,~S.~H.; Shen,~J.; Piecuch,~P. Accurate excited-state
  energetics by a combination of {Monte} {Carlo} sampling and
  equation-of-motion coupled-cluster computations. \emph{The Journal of
  Chemical Physics} \textbf{2019}, \emph{150}, 111101\relax
\mciteBstWouldAddEndPuncttrue
\mciteSetBstMidEndSepPunct{\mcitedefaultmidpunct}
{\mcitedefaultendpunct}{\mcitedefaultseppunct}\relax
\EndOfBibitem
\bibitem[Shen and Piecuch(2012)Shen, and Piecuch]{shen_combining_2012}
Shen,~J.; Piecuch,~P. Combining active-space coupled-cluster methods with
  moment energy corrections via the {CC}( \textit{{P}} ; \textit{{Q}} )
  methodology, with benchmark calculations for biradical transition states.
  \emph{The Journal of Chemical Physics} \textbf{2012}, \emph{136},
  144104\relax
\mciteBstWouldAddEndPuncttrue
\mciteSetBstMidEndSepPunct{\mcitedefaultmidpunct}
{\mcitedefaultendpunct}{\mcitedefaultseppunct}\relax
\EndOfBibitem
\bibitem[Bauman \latin{et~al.}(2017)Bauman, Shen, and
  Piecuch]{bauman_combining_2017}
Bauman,~N.~P.; Shen,~J.; Piecuch,~P. Combining active-space coupled-cluster
  approaches with moment energy corrections via the {CC}( \textit{{P}} ;
  \textit{{Q}} ) methodology: connected quadruple excitations. \emph{Molecular
  Physics} \textbf{2017}, \emph{115}, 2860--2891\relax
\mciteBstWouldAddEndPuncttrue
\mciteSetBstMidEndSepPunct{\mcitedefaultmidpunct}
{\mcitedefaultendpunct}{\mcitedefaultseppunct}\relax
\EndOfBibitem
\bibitem[Deustua \latin{et~al.}(2017)Deustua, Shen, and
  Piecuch]{deustua_converging_2017}
Deustua,~J.~E.; Shen,~J.; Piecuch,~P. Converging {High}-{Level}
  {Coupled}-{Cluster} {Energetics} by {Monte} {Carlo} {Sampling} and {Moment}
  {Expansions}. \emph{Physical Review Letters} \textbf{2017}, \emph{119}\relax
\mciteBstWouldAddEndPuncttrue
\mciteSetBstMidEndSepPunct{\mcitedefaultmidpunct}
{\mcitedefaultendpunct}{\mcitedefaultseppunct}\relax
\EndOfBibitem
\bibitem[Booth \latin{et~al.}(2013)Booth, Gr{\"u}neis, Kresse, and
  Alavi]{booth_towards_2013}
Booth,~G.~H.; Gr{\"u}neis,~A.; Kresse,~G.; Alavi,~A. Towards an exact
  description of electronic wavefunctions in real solids. \emph{Nature}
  \textbf{2013}, \emph{493}, 365--370\relax
\mciteBstWouldAddEndPuncttrue
\mciteSetBstMidEndSepPunct{\mcitedefaultmidpunct}
{\mcitedefaultendpunct}{\mcitedefaultseppunct}\relax
\EndOfBibitem
\bibitem[Cleland(2009)]{cleland_initiator_2009}
Cleland,~D. The initiator {Full} {Configuration} {Interaction} {Quantum}
  {Monte} {Carlo} method: {Development} and applications to molecular systems.
  Ph.D.\ thesis, University of Cambridge, 2009\relax
\mciteBstWouldAddEndPuncttrue
\mciteSetBstMidEndSepPunct{\mcitedefaultmidpunct}
{\mcitedefaultendpunct}{\mcitedefaultseppunct}\relax
\EndOfBibitem
\bibitem[Spencer \latin{et~al.}(2019)Spencer, Blunt, Choi, Etrych, Filip,
  Foulkes, Franklin, Handley, Malone, Neufeld, Di~Remigio, Rogers, Scott,
  Shepherd, Vigor, Weston, Xu, and Thom]{spencer_hande-qmc_2019}
Spencer,~J.~S.; Blunt,~N.~S.; Choi,~S.; Etrych,~J.; Filip,~M.-A.; Foulkes,~W.
  M.~C.; Franklin,~R. S.~T.; Handley,~W.~J.; Malone,~F.~D.; Neufeld,~V.~A.;
  Di~Remigio,~R.; Rogers,~T.~W.; Scott,~C. J.~C.; Shepherd,~J.~J.;
  Vigor,~W.~A.; Weston,~J.; Xu,~R.; Thom,~A. J.~W. The {HANDE}-{QMC} {Project}:
  {Open}-{Source} {Stochastic} {Quantum} {Chemistry} from the {Ground} {State}
  {Up}. \emph{Journal of Chemical Theory and Computation} \textbf{2019},
  \emph{15}, 1728--1742\relax
\mciteBstWouldAddEndPuncttrue
\mciteSetBstMidEndSepPunct{\mcitedefaultmidpunct}
{\mcitedefaultendpunct}{\mcitedefaultseppunct}\relax
\EndOfBibitem
\bibitem[Sun \latin{et~al.}(2018)Sun, Berkelbach, Blunt, Booth, Guo, Li, Liu,
  McClain, Sayfutyarova, Sharma, Wouters, and Chan]{sun_pyscf:_2018}
Sun,~Q.; Berkelbach,~T.~C.; Blunt,~N.~S.; Booth,~G.~H.; Guo,~S.; Li,~Z.;
  Liu,~J.; McClain,~J.~D.; Sayfutyarova,~E.~R.; Sharma,~S.; Wouters,~S.;
  Chan,~G. K.-L. PySCF: the Python-based simulations of chemistry framework.
  \emph{Wiley Interdiscip. Rev.: Comput. Mol. Sci.} \textbf{2018}, \emph{8},
  e1340\relax
\mciteBstWouldAddEndPuncttrue
\mciteSetBstMidEndSepPunct{\mcitedefaultmidpunct}
{\mcitedefaultendpunct}{\mcitedefaultseppunct}\relax
\EndOfBibitem
\bibitem[Werner \latin{et~al.}(2012)Werner, Knowles, Knizia, Manby, and
  Sch{\"u}tz]{MOLPRO-WIREs}
Werner,~H.-J.; Knowles,~P.~J.; Knizia,~G.; Manby,~F.~R.; Sch{\"u}tz,~M.
  {Molpro: a general-purpose quantum chemistry program package}. \emph{WIREs
  Comput Mol Sci} \textbf{2012}, \emph{2}, 242--253\relax
\mciteBstWouldAddEndPuncttrue
\mciteSetBstMidEndSepPunct{\mcitedefaultmidpunct}
{\mcitedefaultendpunct}{\mcitedefaultseppunct}\relax
\EndOfBibitem
\bibitem[Poole(2015)]{tpoole_2015}
Poole,~T. Calculating derivatives within quantum Monte Carlo. Ph.D.\ thesis,
  Imperial College London, 2015\relax
\mciteBstWouldAddEndPuncttrue
\mciteSetBstMidEndSepPunct{\mcitedefaultmidpunct}
{\mcitedefaultendpunct}{\mcitedefaultseppunct}\relax
\EndOfBibitem
\bibitem[Cleland \latin{et~al.}(2012)Cleland, Booth, Overy, and
  Alavi]{cleland_taming_2012}
Cleland,~D.; Booth,~G.~H.; Overy,~C.; Alavi,~A. Taming the {First}-{Row}
  {Diatomics}: {A} {Full} {Configuration} {Interaction} {Quantum} {Monte}
  {Carlo} {Study}. \emph{Journal of Chemical Theory and Computation}
  \textbf{2012}, \emph{8}, 4138--4152\relax
\mciteBstWouldAddEndPuncttrue
\mciteSetBstMidEndSepPunct{\mcitedefaultmidpunct}
{\mcitedefaultendpunct}{\mcitedefaultseppunct}\relax
\EndOfBibitem
\end{mcitethebibliography}

\providecommand{\latin}[1]{#1}
\makeatletter
\providecommand{\doi}
  {\begingroup\let\do\@makeother\dospecials
  \catcode`\{=1 \catcode`\}=2 \doi@aux}
\providecommand{\doi@aux}[1]{\endgroup\texttt{#1}}
\makeatother
\providecommand*\mcitethebibliography{\thebibliography}
\csname @ifundefined\endcsname{endmcitethebibliography}
  {\let\endmcitethebibliography\endthebibliography}{}

 \end{document}